%% file: ms.tex
\documentclass[twocolumn]{IEEEtran}

\input{Def}

\newlength\myindent
\newtheorem{prop}{Proposition}

\title{Structured Covariance Matrix Estimation with Missing-Data for Radar Applications via Expectation–Maximization}

\author{Augusto Aubry, \emph{Senior Member, IEEE}, Antonio De Maio, \emph{Fellow, IEEE}, Stefano Marano, \emph{Senior Member, IEEE}, and Massimo Rosamilia, \emph{Student Member, IEEE}
\thanks{Augusto Aubry, Antonio De Maio (Corresponding Author) and Massimo Rosamilia are with Universit\`a degli Studi di Napoli ``Federico II'', DIETI, Via Claudio 21, I-80125 Napoli, Italy. E-mail: augusto.aubry@unina.it, ademaio@unina.it, massimo.rosamilia@unina.it.}
\thanks{Stefano Marano is with the Dipartimento di Ingegneria dell’Informazione ed Elettrica e Matematica Applicata, University of Salerno, Fisciano (SA) I-84084, Italy.  E-mail: marano@unisa.it.}
}

\begin{document}
\maketitle

\begin{abstract}
Structured covariance matrix estimation in the presence of missing data is addressed in this paper
with emphasis on radar signal processing applications. After a motivation of the study, the array
model is specified and the problem of computing the maximum likelihood estimate of a structured covariance matrix is formulated. A general procedure to optimize the observed-data likelihood function is developed resorting to the expectation-maximization algorithm. 
The corresponding convergence properties are thoroughly established and the rate of convergence is analyzed. The estimation technique is contextualized for two practically relevant radar problems: beamforming and detection of the number of sources. In the former case an adaptive beamformer leveraging the EM-based estimator is presented; in the latter, detection techniques generalizing the classic Akaike information criterion, minimum description length, and Hannan–Quinn information criterion, are introduced. Numerical results are finally presented to corroborate the theoretical study.
\end{abstract}

\section{Introduction}
Missing sensor measurements can arise in a variety of radar signal processing problems as for instance beamforming, direction of arrival estimation, interference cancellation, covariance estimation, and target detection. This demands the development of specific procedures in order to keep contained the loss with respect to the benchmark case where measurements are available from all the sensors. Some practical and interesting situations which are part of this challenging layout are described in the following.
\begin{itemize}
     \item Distributed radar systems (DRSs)~\cite{4102876} which consist of a multitude of stand-alone coherent radar modules (radar satellites). These nodes are equipped with a transmitter, a receiver, an antenna element, and a small processor that conveys the raw data from the receiver via an Internet connection (cable, fiber or wireless) to the beamformer fusion center. Wirelessly networked aperstructure digital phased array radar (WNADPAR)~\cite{WNADPARThesis} represents a perfect example with reference to ship-based surveillance. Therein, the radar satellites are distributed all over the available areas of the ship surface to improve the radar detection range and form narrow beams. Missing observations may arise due to transmission-reception failures.

    \item Switched array~\cite{1993DeLong, 8728188, 5494469, 5352421, 6955686} applications when the number of available channels is smaller than the number of sub-array (or array elements). The system can randomly choose the channel connections on a snapshot-to-snapshot (possibly a block-snapshot to block-snapshot) base or use a pre-fixed switching scheme. From a mathematical point of view, this is tantamount to puncturing dynamically the entire array output vector to get a reduced-size observation vector.

    \item Intermittent sensor failures which can occur due to saturation of some sub-arrays (notice that they can exhibit different antenna patterns and hence can experience different saturation conditions), impulsive noise bursts originated within the channels of failing sensors~\cite{5422700}, and malfunctions at the analog-to-digital converter (ADC) level~\cite[pg. 32]{tuzlukov2017signal}. Actually, failure modes, effects, and criticality analysis (FMECA) on array system level faults reports cases of intermittent troubles, probably caused by material-interaction (i.e., package moulding contamination, surface-state effects, etc.), stress (i.e., burnout, electro-migration, etc.), mechanical (i.e., solder joint failure, die fracture, etc.) and environmental impairments (i.e., temperature, humidity and hydrogen effects)~\cite{WILEMAN201356}. Besides, random failure of array elements has also been discussed in~\cite{8765381, 5960622}.
\end{itemize}
Motivated by the above discussion, this paper deals with the problem of estimating a structured covariance matrix under the missing-data context. 

First of all, the homogeneous Gaussian environment observation model with missing-data is introduced capitalizing on a-priori information about the covariance matrix structure and/or specific array configurations.
Then, the covariance matrix estimation process is formulated as an appropriately constrained optimization problem which is in general difficult to solve. Hence, an effective iterative solution technique, based on the expectation-maximization (EM) algorithm~\cite{10.2307/2984875, vantrees4, 10.2307/1165260, 10.2307/2240463}, is introduced together with some convergence properties and rate of convergence results. 
Each iteration of the algorithm, for very common covariance structures of practical interest in radar signal processing applications, involves only closed form solutions for the unknowns.
It is worth pointing out that the EM algorithm has been already applied to some covariance estimation applications within the statistical literature~\cite{8648152, 9238471, LIU2019278} and in some radar signal processing contexts~\cite{1399075, WANG2005191}. In this respect the main contribution of this paper relies on the inclusion of the constrained case and the analysis of the corresponding convergence properties. Besides, the theory is contextualized for a beamforming application and for the problem of detecting the number of sources. 
The study has lead to some efficient methods capable of operating in the presence of missing-data with satisfactory performance.

Even if the paper is focused on spatial processing, the methodology could be imported in a  temporal processing background where some pulses of the received train may experience unwanted sporadic radio frequency interference (RFI). This means that some slow time samples from some given range cells are missed and the lack of this data has to be properly accounted for at the signal processing design stage.

The paper is organized as follows. Section II introduces the system model and defines some covariance matrix uncertainty sets of practical relevance. Section III formulates the structured covariance matrix estimation problem {in the presence of} missing-data and presents tailored iterative solution methods leveraging possible a-priori structural information. Besides, it addresses convergence issues about the proposed techniques.
In Section IV, the performance of the estimators is analyzed in the context of adaptive beamforming and detection of number of sources.
Finally, Section V draws some conclusions and highlights possible future research avenues.

\subsection{Notation}
Boldface is used for vectors $\ba$ (lower case), and matrices $\bA$ (upper case). The $(k, l)$-entry (or $l$-entry) of a generic matrix $\bA$ (or vector $\ba$) is indicated as $\bA(k, l)$ (or $\ba(l)$). $\bI$ and ${\bzero}$ denote respectively the identity matrix and the matrix with zero entries (their size is determined from the context). The all-ones column vector of size $N$ is denoted by $\bone_N$, whereas $\be_k$ denotes the $k$-th column vector of $\bI$, whose size is determined from the context. Besides, $\diag({\bm{x}})$ indicates the diagonal matrix whose $i$-th diagonal element is $\bx(i)$. The transpose, the conjugate, and the conjugate transpose operators are denoted by the symbols $(\cdot)^{\mathrm{T}}$, $(\cdot)^{*}$, and $(\cdot)^\dagger$, respectively. The determinant, the trace, and the rank of the matrix ${\bm{A}} \in \mathbb{C}^{N\times N}$ are indicated with $\det \left( {\bm{A}} \right)$,  $\tr\{\bm{A}\}$, {and $\mbox{Rank}(\bA)$}, respectively.
$\mathbb{R}^N$ and ${\mathbb{C}}^N$ are respectively the sets of $N$-dimensional column vectors of real and complex numbers. ${\mathbb{H}}^N$ and ${\mathbb{H}}^{N}_{++}$ represent the set of $N\times N$ Hermitian matrices and  Hermitian positive definite matrices, respectively. {Moreover, $\mathbb{R^{++}}$ denotes the set of real numbers greater than zero.} The curled inequality symbol $\succeq$ (and its strict form $\succ$) is used to denote generalized matrix inequality: for any $\bA\in{\mathbb{H}}^N$, $\bA\succeq\bzero$ means that $\bA$ is a positive semi-definite matrix ($\bA\succ\bzero$ for positive definiteness). Besides, for any $\bA\in{\mathbb{H}}^N$, $\lambda_\text{max}(\bA)$ and $\lambda_\text{min}(\bA)$ are the maximum and the minimum eigenvalue of $\bA$, respectively. The letter $j$ represents the imaginary unit (i.e., $j=\sqrt{-1}$). For any complex number $x$, $|x|$ indicates the modulus of $x$. Moreover, for any $\bx \in \mathbb{C}^N$, $\|\bx\|$ denotes the Euclidean norm{, whereas the Frobenius norm of a matrix $\bA$ is indicated as $\|\bA\|_F$.} Let $f(\bx) \in \mathbb{R}$ be a real-valued function, $\nabla_{\bx} f(\bx)$ denotes the gradient of $f(\cdot)$ with respect to $\bx$, with the partial derivatives arranged in a column vector. Furthermore, for any $x, y \in \mathbb{R}$, $\max(x, y)$ returns the maximum between the two argument values. Finally, $\mathbb{E}[\cdot]$ stands for statistical expectation.

\section{Problem Formulation}\label{section:probl_form}
Let us consider a radar system that collects spatial data via a narrow-band array composed of $N$ antennas {and operating in the presence of} noise and interference, with unknown spectral characteristics. Let us suppose that a set of spatial snapshots $\br_i, \; i = 1,\dots, K$, modeled as independent and identically distributed (IID) zero-mean circularly symmetric Gaussian random vectors {(homogeneous environment)} with unknown but structured covariance matrix, is available. Specifically
\begin{equation}
    \br_i \sim CN(0, \bM), \; \bM \in \mathcal{C} \subseteq {\mathbb{H}}^{N}_{++}, \; i=1,\dots,K ,
\end{equation}
where $\mathcal{C}$ denotes the subset of covariance matrices that can generate the observables. Enforcing $\bM$ to belong to $\mathcal{C}$ is tantamount to exploiting some problem structure stemming from a-priori knowledge about the operating environment and/or the array configuration. 
Some practical examples of covariance matrix uncertainty sets are now illustrated.
\begin{enumerate}
    \item Structured covariance matrix with a lower bound on the white disturbance power level~\cite{892662, 6558039}
\begin{equation}\label{set:cm_lb_white_dist_pwr_level}
    \mathcal{C} = \left\{\begin{matrix}
\bM = \sigma_n^2 \bI + \bR_e \\ 
\bR_e \succeq \bzero \\ 
\sigma_n^2 \ge \sigma^2
\end{matrix}\right. ,
\end{equation}
where $\bR_e$ accounts for colored interference and clutter, $\sigma_n^2 > 0$ is the power of the white disturbance term, and $\sigma^2 > 0$ is a known lower bound on the white disturbance power.

\item Structured covariance matrix with a condition number constraint~\cite{6166345}
\begin{equation}
    \mathcal{C} = \left\{\begin{matrix}
    \bM = \sigma_n^2 \bI + \bR_e \\ 
    \bR_e \succeq \bzero \\ 
    \sigma_n^2 \ge \sigma^2 \\
    \frac{\lambda_\text{max}(\bM)}{\lambda_\text{min}(\bM)} \le K_{max}
    \end{matrix}\right. ,
\end{equation}
where $\bR_e$, $\sigma_n^2$, and $\sigma^2$ are defined as in~(\ref{set:cm_lb_white_dist_pwr_level}),  whereas $K_{max}$ is an upper bound to the covariance condition number.

\item {Structured covariance matrix with a rank constraint and a lower bound on the white disturbance power level}~\cite{6809931}
\begin{equation}
    \mathcal{C} = \left\{\begin{matrix}
\bM = \sigma_n^2 \bI + \bR_e \\ 
\bR_e \succeq \bzero \\ 
\mbox{Rank}(\bR_e) \le r \\
\sigma_n^2 \ge \sigma^2
\end{matrix}\right. ,
\end{equation}
where $\bR_e$, $\sigma_n^2$, {and $\sigma^2$} are defined as in~(\ref{set:cm_lb_white_dist_pwr_level}), whereas $r$ is the maximum rank of $\bR_e$.

\item {Structured covariance matrix with a rank constraint}~\cite{vantrees4, 206185}
\begin{equation}\label{set:fixed_rank}
    \mathcal{C} = \left\{\begin{matrix}
\bM = \sigma_n^2 \bI + \bV\bS_f \bV^\dagger \\ 
\bV\bS_f \bV^\dagger \succeq \bzero \\ 
\mbox{Rank}(\bV\bS_f \bV^\dagger) \le d \\
{\sigma_n^2 > 0}
\end{matrix}\right. ,
\end{equation}
where {$d$ is an upper bound to $\mbox{Rank}(\bV\bS_f \bV^\dagger)$}, $\sigma_n^2$ is defined as in~(\ref{set:cm_lb_white_dist_pwr_level}), $\bV$ is an $N \times d$ array manifold matrix {(which can be modeled either as a known or as an unknown parameter)}, $\bS_f$ denotes the  $d \times d$ diagonal sources covariance matrix, whereas $d \le N$ is the number of sources.

\item Structured covariance matrix with a centro-Hermitian symmetry~\cite{1093391}
\begin{equation}\label{set:persymmetry}
   \mathcal{C} = \left\{\begin{matrix}
\bM = \bJ {\bM}^{*} \bJ \\ 
\bM \succeq \bzero \\
\end{matrix}\right. ,
\end{equation}
with $\bJ$ the $N\times N$ permutation matrix given by
\begin{equation}\label{eq:J_exchange_matrix}
    \bJ = \begin{bmatrix}
        0 & 0 & \cdots & 0 & 1 \\ 
        0 & 0 & \cdots & 1 & 0\\ 
        \vdots & \vdots & \ddots & \vdots & \vdots\\ 
        1 & 0 & \cdots & 0 & 0
        \end{bmatrix}
\end{equation}
\end{enumerate}

Besides, any combination of the above uncertainty sets (corresponding to their intersection) constitutes additional interesting examples.

The estimation problem (object of the present paper) demands a data model endowed with the capability to handle missing-data arising from the lack of some entries within specific spatial snapshots.
To this end, each observed snapshot is modeled as
\begin{equation}
    \by_i = \bA_i \br_i,  \; i=1,\dots,K ,
\end{equation}
where $\bA_i$ is the $p_i\times N$ selection matrix, constructed by extracting from $\bI$ the $p_i \le N$ rows corresponding to the available observations at the $i$-th snapshot.
In the following, the vectors $\br_i$ and $\by_i$ will be referred to as \textit{complete} and \textit{observed} data, respectively. 

\section{Covariance Matrix Estimation Procedure}
This section is devoted to the derivation of a covariance matrix estimation procedure in the presence of missing-data accounting for model structures via suitable constraints.
As already pointed out, the problem is of primary importance for many applications in the field of radar signal processing~\cite{1100705, 31267, LINEBARGER199585, li2005robust, 937465, 1166614, 890324, 5514423} and, in most cases, a maximum likelihood (ML) estimator is usually demanded at least due to its favorable asymptotic properties. For the missing-data case, the constrained ML estimate of the covariance matrix, given the observed-data, can be formulated as 
\begin{equation} \label{eq:problem}
\hat{\bM}(\btheta) = \operatorname*{arg\,max}\limits_{\bM(\btheta) \in \mathcal{C}} \mathcal{L}_y(\bM(\btheta) | \bY, \bA_1, \dots, \bA_K)  ,
\end{equation}
with
\begin{equation}
		\begin{aligned}
	\label{eq:observed_log_likelihood}
		& \mathcal{L}_y(\bM(\btheta) | \bY, \bA_1, \dots, \bA_K) =& \\
		&\qquad  - \sum_{i=1}^K p_i  \ln(\pi) - \sum_{i=1}^K \ln(\det(\bA_i \bM(\btheta) \bA_i^\dagger)) +& \\ 
		 &\quad \quad \quad \qquad \qquad \qquad +\tr\{(\bA_i \bM(\btheta)\bA_i^\dagger)^{-1} \by_i \by_i^\dagger\}&
		 \end{aligned}
\end{equation}
the observed-data log-likelihood, $\bY = \{\by_1, \dots, \by_K\}$ the set of observed-data, and $\btheta \in \mathbb{R}^{V}$ the vector of the unknown parameters defining the underlying structure of $\bm{M}$. This is tantamount to solving
\begin{equation} \label{eq:problem_theta}
\hat{\btheta}_{ML} = \operatorname*{arg\,max}\limits_{\btheta: \bM(\btheta) \in \mathcal{C}} \mathcal{L}_y(\btheta | \bY, \bA_1, \dots, \bA_K).
\end{equation}

Computing $\hat{\btheta}_{ML}$ (or equivalently $\hat{\bM}(\btheta)$) is, in general, a difficult problem for which an analytic closed-form solution could not be available~\cite{1456695}. Besides, an optimization procedure based on a multi-dimensional grid search (MDGS) in the unknown parameter space could be computationally prohibitive. This motivates the interest toward iterative approximated procedures characterized by a more affordable computational cost than ML evaluation via MDGS.

\subsection{EM Algorithm}
EM is a widely adopted iterative technique to obtain approximate ML estimates of parameters
from incomplete-data\footnote{In situations where direct access to the \emph{complete} set of observations is not available, part of the data are missing or, more in general, data undergo a many-to-one mapping before becoming available to the observer.}~\cite{10.2307/2984875, vantrees4, 10.2307/1165260}. 
The algorithm is composed of two steps. In the former, referred to as expectation (E) step, the conditional expectation of the complete-data likelihood, given the observed-data and the current estimate of the parameters, is evaluated (E-step score function). In the latter, referred to as the maximization (M) step, the E-step score function (corresponding to current estimate of the parameters) is maximized with respect to the unknowns.
The EM starts with an initial guess of the parameters, i.e., $\btheta^{(0)}$, and iterates between E and M steps. The procedure can also be interpreted as a minorization-maximization optimization technique where the surrogate function stems from the Jensen inequality~\cite{doi:10.1198/0003130042836}.
With reference to the estimation problem in~\eqref{eq:problem_theta}, at the $h$-th iteration, the E-step consists in the evaluation of the score function
\begin{equation}\label{eq:Q}
    Q\left(\btheta, \btheta^{(h-1)}\right) = \mathbb{E}[\mathcal{L}_r(\btheta) | \bY, \bA_1, \dots, \bA_K, \hat{\bM}(\btheta^{(h-1)})] ,
\end{equation}
where 
\begin{equation}
\begin{aligned}\label{eq:complete_log_likelihood}
    \mathcal{L}_r(\btheta) =& -K[N \ln(\pi) + \ln(\det(\bM(\btheta))) \\ 
    & + \tr\{\bM(\btheta)^{-1} \bS\}]
\end{aligned}
\end{equation}
is the complete-data log-likelihood, 
\begin{equation}\label{eq:sample_covariance}
    \bS = \frac{1}{K} \sum_{i=1}^K \br_i \br_i^\dagger
\end{equation}
is the sample covariance matrix of the complete-data, and $\hat{\bM}(\btheta^{(h-1)})$ is the estimate of the covariance matrix at the $(h-1)$-iteration.
Computing the conditional expectation involved in~\eqref{eq:Q} yields
\begin{equation}\label{eq_E_step}
    \begin{aligned}
    Q(\btheta, \btheta^{(h-1)}) =&-K[N \ln(\pi) + \ln(\det(\bM(\btheta)))  \\ 
    & + \tr\{\bM(\btheta)^{-1} \bSigma^{(h-1)}\}], 
\end{aligned}
\end{equation}
where
\begin{equation}\label{eq:sample_covariance_EM_obs_data}
    \bSigma^{(h-1)} = \frac{1}{K} \sum_{i=1}^K \bC_i^{(h-1)}
\end{equation}
{is} the sample mean of the conditional correlation matrices
\begin{equation}\label{eq:cond_expectation_incomplete_data}
    \bC_i^{(h-1)} = \mathbb{E}[\br_i \br_i^\dagger | \by_i, \bA_i, \hat{\bM}(\btheta^{(h-1)})], \; i=1,\dots, K.
\end{equation}

A closed-form expression to
\begin{equation}
    \mathbb{E}[\br_i \br_i^\dagger |\by_i, \bA_i, \bM] = \bC_i ,
\end{equation}
is given by {(see Appendix A for the detailed derivation)}
\begin{equation}\label{eq:C_i}
\begin{aligned}
    \bC_i = & (\bA_i^\dagger \by_i + \bar{\bA_i}^\dagger \bmu_i) (\bA_i^\dagger \by_i + \bar{\bA_i}^\dagger \bmu_i)^\dagger + \bar{\bA_i}^\dagger \bG_i \bar{\bA_i}
\end{aligned}
\end{equation}
with $\bar{\bA}_i$ the $N-p_i\times N$ {selection matrix} complementary to $\bA_i$ (obtained removing from $\bI$ the $p_i$ rows corresponding to $\bA_i$),
\begin{equation}
    \bmu_i =  \bar{\bA}_i \bM \bA_i^\dagger (\bA_i \bM \bA_i^\dagger)^{-1} \by_i 
\end{equation}
and
\begin{equation}
\begin{aligned}
       & \bG_i = \bar{\bA}_i \bM \bar{\bA}_i^\dagger -  \bar{\bA}_i \bM \bA_i^\dagger (\bA_i \bM \bA_i^\dagger)^{-1} \bA_i \bM \bar{\bA}_i^\dagger    .
\end{aligned}  
\end{equation}

After an E-step, an M-step is performed, corresponding to the maximization of the score function~\eqref{eq_E_step}, namely the estimate of the parameters is updated according to
\begin{equation}\label{eq:Mstep}
 \btheta^{(h)} = \operatorname*{arg max}\limits_{\btheta:\bM(\btheta) \in \mathcal{C}} Q(\btheta, \btheta^{(h-1)}) .
\end{equation}
The following proposition {outlines} the main features of the sequence of estimates.

\begin{prop}\label{proposition:1} 
	Provided that $\bM(\btheta^{(0)})\succ \bf 0$, $K\geq N$, and $\mathcal{C} = \mathcal{B} \cap {\mathbb{H}}_{++}^N$, with $\mathcal{B}$ a closed set of ${\mathbb{H}}^N$, then
	\begin{itemize}
		\item $\bM(\btheta^{(h)})\succ0$, for all $h \ge 0$ and $\mathcal{L}_y(\bM(\btheta^{(h)}) | \bY, \bA_1, \dots, \bA_K)$ is a monotonically increasing sequence;
		\item if $\mathcal{B} \in {\mathbb{H}}_{++}^N$ is  a closed set of positive definite matrices, then $\bM(\btheta^{(h)}), \; h\ge0, $ is a bounded sequence and $\mathcal{L}_y(\bM(\btheta^{(h)}) | \bY, \bA_1, \dots, \bA_K)$, $h\ge0$, converges to a finite value. Besides, supposing  $M(\btheta)$ a norm coercive differentiable mapping, any limit point $\btheta^*$ to $\btheta^{(h)}$ is a B-stationary point~\cite{10.1287/moor.2016.0795, 10.2307/25151818, 7736116, 8239836} to Problem~(\ref{eq:problem_theta}).
	\end{itemize}
\end{prop}
\begin{IEEEproof}
	See Appendix B.
\end{IEEEproof}

A summary of the EM procedure is reported in~Algorithm~\ref{alg:EM}, where{, leveraging the results of Proposition 1,} the exit condition of the procedure is set as $|{P}^{(h)}- {P}^{(h-1)}| \le \varepsilon_1$ or $\|{\btheta}^{(h)}- {\btheta}^{(h-1)}\| \le \varepsilon_2$, where $\varepsilon_1, \varepsilon_2 > 0$ and
\begin{equation}\label{eq:P_Q}
    P^{(h)} = \mathcal{L}_y(\bM(\btheta^{(h)}) | \bY, \bA_1, \dots, \bA_K).
\end{equation}

\begin{algorithm}
\caption{EM {Covariance Matrix} Estimation Procedure}
\label{alg:EM}
\begin{algorithmic}
\REQUIRE $N$, $K$, $\bY$, $\bA_1, \dots, \bA_K$, $\mathcal{C}$, $\btheta^{(0)}$, $\varepsilon_1$, $\varepsilon_2$.\\
\ENSURE {EM} solution $\hat{\btheta}$ to Problem~(\ref{eq:problem_theta}).
\STATE{\textbf{Initialization}
\STATE $h = 0$;
\STATE  ${P}^{(0)} = \mathcal{L}_y(\bM(\btheta^{(0)}) | \bY, \bA_1, \dots, \bA_K)$};
\STATE \textbf{repeat} 
\begin{enumerate}[label={\theenumi:}]
 \STATE $h = h + 1 $;
 \STATE E-Step: Compute $\bSigma^{(h-1)}$ {given by}~\eqref{eq:sample_covariance_EM_obs_data};
 \STATE M-Step: Find $\btheta^{(h)}$ using \eqref{eq:Mstep};
 \STATE Compute ${P}^{(h)}$ using~\eqref{eq:P_Q};
 \end{enumerate}
 \STATE \textbf{until} {$|{P}^{(h)}- {P}^{(h-1)}| > \varepsilon_1$ or $\|{\btheta}^{(h)}- {\btheta}^{(h-1)}\| > \varepsilon_2$};
 \STATE Output $\hat{\btheta} = \btheta^{(h)}$.
\end{algorithmic}
\end{algorithm}
{\bf Remark 1.} Before proceeding further, a useful digression on the convergence rate of Algorithm~\ref{alg:EM} is now in order. As shown in~\cite{10.2307/2984875}, assuming that ${\btheta}^{(h)}$ converges to the ML estimate {$\hat{\btheta}_{ML}$}, then the rate of convergence is ruled by the spectral radius $\rho(\bR^{EM})$ of the rate matrix
\begin{equation}
\bR^{EM} = \bI - \bF_{obs}^{\frac{1}{2}} \bF_{EM}^{-1} \bF_{obs}^{\frac{1}{2}} ,
\end{equation}
{where 
\begin{equation}\label{eq:fobs}
	\bF_{obs} = \left. - \nabla_{\btheta} \nabla_{\btheta}^{\mathrm{T}}  \mathcal{L}_y(\btheta) \right|_{ \btheta = {\hat{\btheta}_{ML}}}
\end{equation}
is the observed information matrix and 
\begin{equation}\label{eq:fem}
	\bF_{EM} =  \left. \mathbb{E}\left[ - \nabla_{\btheta} \nabla_{\btheta}^{\mathrm{T}}  \mathcal{L}_r(\btheta) | \bY, \btheta \right] \right|_{\btheta = {\hat{\btheta}_{ML}}}
\end{equation}
is the expected complete information matrix.}
See Appendix D for the computation of \eqref{eq:fobs} and \eqref{eq:fem} with reference to \eqref{eq:observed_log_likelihood} and \eqref{eq:complete_log_likelihood}.

Just to provide a study example, let us consider $N=10$ and covariance matrix belonging to~\eqref{set:fixed_rank}. In particular, the true parameters are $\bV =  \be_1$, $\bS_f = 10$, and $\sigma_n^2 = 1$. As to the missing-data {pattern}, the selection matrices obtained skipping the zero rows of
\begin{enumerate}
\item $\diag(\bone_N - \be_1 - \be_3)$,
\item $\diag(\bone_N - \be_2 - \be_5)$,
\item $\diag(\bone_N - \be_4 - \be_7)$,
\item $\diag(\bone_N - \be_6 - \be_8)$,
\item $\diag(\bone_N - \be_9 - \be_{10})$,
\end{enumerate} 
{are cyclically used (according to the reported order) to select the observations at the different snapshots.}

\begin{figure}[htbp!] 
\centering
\label{fig:convergence_rate_3}	\includegraphics[width=0.95\linewidth]{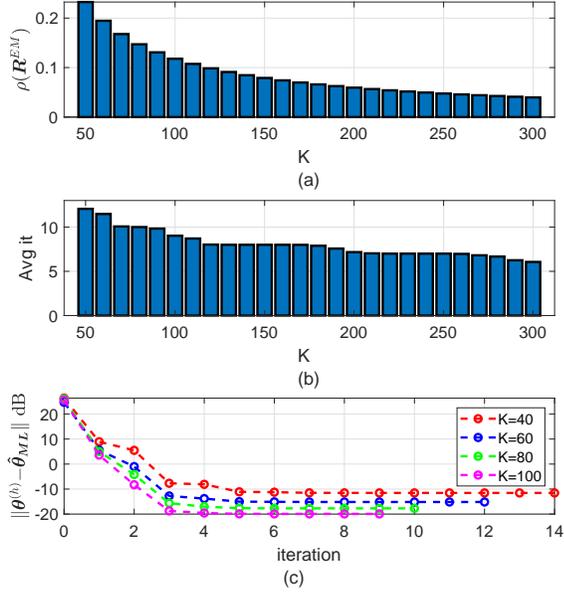}
\caption{Convergence rate analysis for the case study discussed in the main text, with $N=10$. Fig. (a) displays the average rate of convergence versus the number of snapshots, while Fig. (b) displays the average number of iterations versus the number of snapshots. The {norm difference} $\|\btheta^{(h)} - \hat{\btheta}_{ML}\|$ in dB {versus the number of iterations for Algorithm~\ref{alg:EM} is reported in Fig. (c)}, assuming $K=40,60,80,100$.}
\label{fig:convergence_rate} 
\end{figure}

Figs.~\ref{fig:convergence_rate} (a) and (b) display {respectively} the average convergence rate and the average number of iterations, required by Algorithm~\ref{alg:EM} ({with $\varepsilon_1 = \varepsilon_1 = 10^{-7}$ and }initialized using the sample covariance matrix of $2N$ IID zero-mean circularly symmetric Gaussian random vectors) to achieve convergence, versus the number of snapshots. {The {results} are obtained} via standard Monte Carlo counting techniques over $100$ independent trials.
Inspection of the figures outlines that a lower value of $\rho(\bR^{EM})$ is {associated with} a faster convergence of Algorithm~\ref{alg:EM}.

In Fig.~\ref{fig:convergence_rate} (c){, for a given trial, the distance between the ML estimate and the EM solution at the $h$-th M-step, i.e.,} $\|\btheta^{(h)} - \hat{\btheta}_{ML}\|$, is {plotted versus the number of iterations}, assuming $K=40,60,80,100$.
This analysis corroborates that increasing the number of available snapshots a smaller number of iterations is required for Algorithm~\ref{alg:EM} to converge. Furthermore, the larger the number of snapshots the closer the final estimate of Algorithm~\ref{alg:EM} to $\hat{\btheta}_{ML}$.

{\bf Remark 2.}
It is worth pointing out that the main advantage connected with the use of an EM algorithm occurs when the optimization involved in~\eqref{eq:Mstep} is more tractable than the direct maximization of the observed-data likelihood~\eqref{eq:problem_theta}. {It is clear that the crucial point to devise an EM-based constrained covariance estimation procedure is the capability to obtain an optimal solution to~\eqref{eq:Mstep} with {an} {affordable} computational effort}.
Besides, it is important to remark that different system constraints generally induce distinct feasible sets that generally result in different solutions $\btheta^{(h)}$.
In particular, for the special case of unconstrained estimation~\cite{vantrees4},
\begin{equation}\label{eq:sol_M_step_uncontrained}
    \hat{\bM}(\btheta^{(h)}) =  \bSigma^{(h-1)}
\end{equation}
is the maximizer of $Q(\btheta, \btheta^{(h-1)})$ and therefore {it} constitutes the updated estimate $\btheta^{(h)}$. 


In the following subsections, two well-known radar applications are analyzed in the missing-data scenario: adaptive beamforming and detection of the sources number. In particular, each application is presented and the underlying structured covariance matrix estimation problem is discussed. Then, EM-based solution methods, leveraging problem structure at different extents, are devised.
\subsection{Adaptive Beamforming}
Let us consider the minimum variance distortionless response (MVDR) (also known as the Capon) beamformer~\cite{vantrees4}
\begin{equation}\label{eq:Capon}
    \bw = \frac{{\bM}^{-1} \bv(\theta_0)}{\bv(\theta_0)^\dagger {\bM}^{-1} \bv(\theta_0) },
\end{equation}
where $\bv(\theta_0)$ is the steering vector in the direction {of interest} $\theta_0$.

In a practical scenario the covariance matrix must be estimated from the incoming data leading to an adaptive weight vector.
It {is crystal clear} that obtaining an accurate estimate of the unknown interference covariance matrix is a crucial task affecting the performance of the {resulting} adaptive beamformer.
In a typical case where a set of $K\ge N$ secondary data $\{\br_i\}, \;i=1,\dots,K$, is available, the unstructured ML estimate of $\bM$ is given by the sample covariance matrix $\bS$ (often with a diagonal loading), defined as in (\ref{eq:sample_covariance})~\cite{vantrees4}. Therefore, $\bS$ (or possibly a diagonally  loaded version) is employed in place of $\bM$ in~\eqref{eq:Capon}, obtaining the MVDR adaptive beamformer.

Let us now focus on a missing-data context where the problem of computing the ML estimate of the covariance matrix from the observed-data is described in~\eqref{eq:problem} and a viable estimation procedure is reported in Algorithm~\ref{alg:EM}. 
As a consequence, following the same guideline as in the definition of the MVDR adaptive beamformer\footnote{The analysis developed in the following can be also naturally extended to other kinds of beamformers.}, it is possible to gain adaptivity under the missing-data scenario using
\begin{equation}\label{eq:MVDR_EM}
    \bw_{EM} = \frac{\hat{\bM}_{EM}^{-1} \bv(\theta_0)}{\bv(\theta_0)^\dagger \hat{\bM}_{EM}^{-1} \bv(\theta_0) },
\end{equation}
where $\hat{\bM}_{EM}$ denotes the estimate of the covariance matrix obtained via the EM procedure described in Algorithm~\ref{alg:EM}.

As highlighted in the previous subsection, tailored solutions to the M-step of Algorithm~\ref{alg:EM} can be devised under the assumption of $\bM$ belonging to a specific covariance matrix uncertainty set. In this respect, some case studies are discussed in the following.

\subsubsection{Unconstrained Estimation}
The special case of unconstrained estimation has been described in the previous subsection and a solution to the resulting M-step is given by~\eqref{eq:sol_M_step_uncontrained}.

\subsubsection{Constraint on the lower bound of the white noise power level}
The Fast ML (FML) procedure~\cite{892662, 6558039} provides the M-step
solution when $\bM$ belongs to the uncertainty set~(\ref{set:cm_lb_white_dist_pwr_level}). 
Specifically, denoting by $\bU \bLambda_{\bSigma} \bU^\dagger$ the eigenvalue decomposition (EVD) of $\bSigma^{(h-1)}$ and by $\tilde{\lambda}_{v}, \;v=1,\dots, N$ its eigenvalues, at the M-step update under the uncertainty set~(\ref{set:cm_lb_white_dist_pwr_level}), is given by 
\begin{equation}\label{eq:Mstep_FML}
    \hat{\bM}(\btheta^{(h)}) = \bU \bLambda_{FML} \bU^\dagger ,
\end{equation}
with
\begin{equation}
    \bLambda_{FML} = \diag(\lambda_{1, FML}, \dots, \lambda_{N, FML})
\end{equation}
and $\lambda_{v, FML} = \max(\tilde{\lambda}_{v}, \: \sigma^2), \;v=1,\dots, N$.

\subsubsection{Centro-Hermitianity constraint}
In many scenarios of practical interests (standard rectangular, hexagonal, uniform circular or cylindrical array), the covariance matrix exhibits a centro-Hermitian structure~\cite{vantrees4}, which is equivalent to consider $\bM$ belonging to~(\ref{set:persymmetry}).
Capitalizing on the problem structure, the M-step solution can be obtained using the forward-backward (FB) averaged sample covariance matrix procedure~\cite{vantrees4}, resulting into
\begin{equation}\label{eq:Mstep_Persymmetry}
    \hat{\bM}(\btheta^{(h)}) =  \bSigma_{FB} ,
\end{equation}
where
\begin{equation}\label{eq:sigma_FB}
    \bSigma_{FB} = \frac{1}{2} (\bSigma^{(h-1)} + \bJ {\bSigma^{(h-1)}}^{*} \bJ) .
\end{equation}

\subsubsection{Lower bound constraint on the white noise power level plus Centro-Hermitianity}
This is tantamount to considering the uncertainty set characterizing the centro-Hermitian covariance matrices with a lower bound on the white disturbance power level
\begin{equation}
    \mathcal{C} = \left\{\begin{matrix}
\bM = \sigma_n^2 \bI + \bR_e \\ 
\bM = \bJ {\bM}^* \bJ \\ 
\bR_e \succeq \bzero \\ 
\sigma_n^2 \ge \sigma^2
\end{matrix}\right. ,
\label{set2:cm_lb_white_dist_pwr_level_and_persymm}
\end{equation}
where $\bR_e$, $\sigma_n^2$ and $\sigma^2$ are defined as in~(\ref{set:cm_lb_white_dist_pwr_level}), whereas $\bJ$ is given by~(\ref{eq:J_exchange_matrix}).
{In this situation}, denoting by $\bU_{FB} \: \bLambda_{FB} \: \bU_{FB}^\dagger$ the EVD of $\bSigma_{FB}$ {defined in~\eqref{eq:sigma_FB}}, with $\bLambda_{FB} = \diag(\tilde{\lambda}_{1, FB}, \dots, \tilde{\lambda}_{N, FB})$, it follows that the M-step update is now given by
\begin{equation}\label{eq:M_step_FML_persymmetry}
    \hat{\bM}(\btheta^{(h)}) = \bU_{FB} \: \bLambda_{FML-FB} \: \bU_{FB}^\dagger ,
\end{equation}
where
\begin{equation}
    \bLambda_{FML-FB} = \diag(\lambda_{1, FML-FB}, \dots, \lambda_{N, FML-FB}),
\end{equation}
with $\lambda_{v, FML-FB} = \max(\tilde{\lambda}_{v, FB},\: \sigma^2), \;v=1,\dots, N$.

The overall procedure used to find the proposed adaptive Capon beamformer in the context of missing-data is summarized in Algorithm~\ref{alg:adaptive_beamforming}.

\begin{algorithm}
\caption{Adaptive beamforming in the context of missing-data}
\label{alg:adaptive_beamforming}
\textbf{Input:} $N$, $K$, $\bY$, $\bA_1, \dots, \bA_K$, $\mathcal{C}$, $\btheta^{(0)}$, $\varepsilon_1$, $\varepsilon_2$.\\
\textbf{Output:} EM-based adaptive beamformer $\bw_{EM}$.
\begin{enumerate}[label={\theenumi:}]
\item find $\hat{\bM}_{EM}$ via~Algorithm~\ref{alg:EM} using the appropriate {bespoke} solution~\eqref{eq:sol_M_step_uncontrained}, \eqref{eq:Mstep_FML}, \eqref{eq:Mstep_Persymmetry}, \eqref{eq:M_step_FML_persymmetry} to the M-step;
\item compute $\bw_{EM}$ using \eqref{eq:MVDR_EM};
 \item output $\bw_{EM}$ .
\end{enumerate}
\end{algorithm}

\subsection{Detection of Number of Sources}
Let us consider $d$ uncorrelated narrow-band sources impinging the array from distinct directions $\{\theta_s\}, s=1, \dots, d < N$. 
After amplification, down-conversion, and digital sampling, the $i$-th received complete spatial snapshot $\br_i$ is given by
\begin{equation}
    \br_i = \bV \bs_i + \bn_i, \quad i=1,\dots, K
\end{equation}
where $\bV = [\bv(\theta_1), \bv(\theta_2), \dots, \bv(\theta_d)] \in \mathbb{C}^{N\times d}$ is the array manifold matrix (assumed full-rank), {$\bs_i,\; i=1,\dots,K$ are IID zero-mean Gaussian random vectors of sources amplitudes (independent of each other) with powers $\sigma_s^2$,  $s=1,\dots, d$, respectively, and $\bn_i$ are IID zero-mean circularly symmetric Gaussian random vectors with power $\sigma_n^2$, assumed statistically independent from the sources.

For the case at hand, the covariance matrix of the received signal can be assumed belonging to~(\ref{set:fixed_rank}). 
Resorting to the EVD, the complete-data covariance matrix takes on the convenient form
\begin{equation}
    \bM = \sum_{v=1}^d \lambda_v \bPhi_v \bPhi_v^\dagger + \sigma_n^2  \sum_{v=d+1}^N  \bPhi_v \bPhi_v^\dagger ,
\end{equation}
where $\lambda_v, \sigma_n^2$, and ${\bPhi}_v,\; v=1,\dots, N$ denote to the eigenvalues and the corresponding eigenvectors of $\bM$, respectively, {with ${\lambda}_1 \ge {\lambda}_2 \dots {\lambda}_d \ge 0$.}
As a consequence, {denoting by $\bPhi_{v,R}^{\mathrm{T}}$ and $\bPhi_{v,I}^{\mathrm{T}}$ the vectors of the real and imaginary components of $\bPhi_{v}$}, for a given $d$, the vector of the unknown parameters (underlying the covariance structure) is
{\begin{equation}\label{eq:theta_d}
    {\btheta}(d) = [\lambda_1, \dots, \lambda_d, \sigma_n^2, \bPhi_{1,R}^{\mathrm{T}},\bPhi_{1,I}^{\mathrm{T}}, \dots, \bPhi_{d,R}^{\mathrm{T}},  \bPhi_{d,I}^{\mathrm{T}}]^{\mathrm{T}}, 
\end{equation}}
which explicitly reveals the role of $d$ in controlling the degrees of freedom of the covariance matrix.

The approach pursued in this subsection follows from~\cite{vantrees4, 1100705,31267, 10.2307/2985032}, where a source number detection algorithm, based on a data-adaptive test statistic {plus} a penalty function related to the degrees of freedom, is devised. Specifically, denoting by {$\hat{\btheta}_{ML}(d) = [\hat{\lambda}_1, \dots, \hat{\lambda}_d, \hat{\sigma}_n^2, \hat{\bPhi}_{1,R}^{\mathrm{T}},\hat{\bPhi}_{1,I}^{\mathrm{T}}, \dots, \hat{\bPhi}_{d,R}^{\mathrm{T}},  \hat{\bPhi}_{d,I}^{\mathrm{T}}]^{\mathrm{T}}$} the ML estimate of ${\btheta}(d)$, the problem of detecting the number of sources can be formulated as
\begin{equation}\label{eq:find_d_problem_max}
    \hat{d} = \operatorname*{argmax}\limits_{d = 0, \dots, K_1} \mathcal{L}_r(\hat{\btheta}_{ML}(d)) - T(d) , 
\end{equation}
where $K_1 \le N-1$ is an upper bound to the number of sources, $\mathcal{L}_r(\hat{\btheta}_{ML}(d))$ is the statistic given by the complete-data log-likelihood (\ref{eq:complete_log_likelihood}) evaluated at $\hat{\btheta}_{ML}(d)$, and $T(d)$ is a penalty term accounting for the number of free parameters in the assumed model.
In particular, taking the negative value of $\mathcal{L}_r(\hat{\btheta}_{ML}(d))$ and dropping the terms functionally independent from $d$, the following decision statistic is obtained~\cite{anderson1963}
\begin{equation}\label{eq:L_d_statistic}
    L(d, \hat{\lambda}_1, \dots, \hat{\lambda}_N) = K(N-d)\ln{\left\{\frac{ \frac{1}{N-d}\sum_{v=d+1}^{N} \hat{\lambda}_v  }{(\prod_{v=d+1}^{N} \hat{\lambda}_v)^{\frac{1}{N-d}}}\right\}} .
\end{equation}
Exploiting the above result, problem (\ref{eq:find_d_problem_max}) is equivalently recast as\footnote{It is also worth pointing out that~\eqref{eq:find_d_problem_min} can be generalized to the case of covariance matrices with additional structured constraints.}
\begin{equation}\label{eq:find_d_problem_min}
    \hat{d} = \operatorname*{argmin}\limits_{d = 0, \dots, K_1}  L(d, \hat{\lambda}_1, \dots, \hat{\lambda}_N) + p(d),
\end{equation}
where $p(d)$ is a specific penalty function. In the following, three detection tests, Akaike information criterion (AIC)~\cite{1100705}, minimum description length (MDL)~\cite{31267}, and Hannan–Quinn information criterion (HQC)~\cite{10.2307/2985032}, are considered. Each test is characterized by a different penalty function $p(d)$; in particular
\begin{equation}\label{eq:p_d}
p(d) = 
    \left\{\begin{matrix*}[l]
    d(2N - d), & \text{AIC}\\
    1/2\;[d(2N - d) + 1] \ln{K}, & \text{MDL}\\
    [d(2N - d)+1] \ln(\ln(K)) & \text{HQC}
\end{matrix*}\right. .
\end{equation}

Let us now frame the decision statistic in missing-data context.
Accordingly, the criterion~\eqref{eq:find_d_problem_max} can be modified as:
\begin{equation}\label{eq:criter_det_obs}
    \hat{d} = \operatorname*{argmax}\limits_{d = 0, \dots, K_1}
    \mathcal{L}_y(\hat{\btheta}_{ML}(d) | \bY, \bA_1, \dots, \bA_K) - T(d) .
\end{equation}
This requires, for a given $d$, the computation of the ML estimate $\hat{\btheta}_{ML}(d)$ from the observed-data.
To this end, a viable technique is represented by Algorithm~\ref{alg:EM} applied to a covariance uncertainty set including the fixed rank constraint in~(\ref{set:fixed_rank}). Two relevant case studies are thus developed in the following, providing tailored solutions to the M-step.

\subsubsection{Fixed rank constraint}
Let us exploit the knowledge that $\bM$ belongs to~(\ref{set:fixed_rank}). Specifically, for a given $d$, the M-step at the $h$-th iteration is cast as
\begin{equation}\label{eq:problem_m_constrained}
    {\btheta_d}^{(h)} = \operatorname*{argmax}\limits_{\btheta_d} Q\left(\btheta_d, {\btheta_d}^{(h-1)}\right) ,
\end{equation}
where $\btheta_d$ is defined as in~(\ref{eq:theta_d}).
The maximizer of Problem~(\ref{eq:problem_m_constrained}) is given by~\cite{vantrees4}
\begin{equation}\label{eq:solution_str_detection_mstep}
        {\btheta_d}^{(h)} = [\tilde{\lambda}_1, \dots, \tilde{\lambda}_d, \tilde{\sigma}_{n}^2, \tilde{\bPhi}_{1,R}, \tilde{\bPhi}_{1,I}, \dots, \tilde{\bPhi}_{d,R}, \tilde{\bPhi}_{d,I} ]^{\mathrm{T}}, 
\end{equation}
where $\tilde{\lambda}_v$ and $\tilde{\bPhi}_{v},\; v=1,\dots, d$ are the $d$ greatest eigenvalues and the corresponding eigenvectors of $\bSigma^{(h-1)}$, with $\tilde{\bPhi}_{v,R}$, and $\tilde{\bPhi}_{v,I}$ the real and imaginary components of $\tilde{\bPhi}_{v}$, whereas 
\begin{equation}
    \tilde{\sigma}_{n}^2 = \frac{1}{N-d} \sum_{v=N-d}^N \tilde{\lambda}_v
\end{equation}
is the arithmetic mean of the $N - d$ smallest eigenvalues of $\bSigma^{(h-1)}$.

Exploiting the above results, the $h$-th estimate of the covariance matrix is given by
\begin{equation}\label{eq:m_est_det_num_src_str_set}
    \hat{\bM}({\btheta_d}^{(h)}) = \bU \bLambda_S \bU^\dagger + \tilde{\sigma}_{n}^2  \; \bI ,
\end{equation}
where
\begin{equation}
    \bLambda_S = \diag(\tilde{\lambda}_1 - \tilde{\sigma}_{n}^2, \dots, \tilde{\lambda}_d - \tilde{\sigma}_{n}^2), 
\end{equation}
and
\begin{equation}
    \bU = [\tilde{\bPhi}_1, \dots, \tilde{\bPhi}_d].
\end{equation}

Hence, taking the negative value and dropping the constant terms of the observed-data log-likelihood, the order estimate is given by
\begin{equation}\label{eq:det_str_test_Ly}
    \hat{d}_{EM} = \operatorname*{argmin}\limits_{d = 0, \dots, K_1}  L_y({\hat{\btheta}_d}) + p(d) ,
\end{equation}
 where 
 \begin{equation}\label{eq:str_detection_decision_stat_Ly}
\begin{aligned}
     L_y({\hat{\btheta}_d}) = \sum_{i=1}^{K} & \ln(\det(\bA_i \hat{\bM}({\hat{\btheta}_d})\bA_i^\dagger )) \\ & + \tr\{(\bA_i \hat{\bM}({\hat{\btheta}_d})\bA_i^\dagger)^{-1} \by_i \by_i^\dagger\}
    \end{aligned}
\end{equation}
with ${\hat{\btheta}_d}$ the final estimate of Algorithm~\ref{alg:EM} and $p(d)$ a specific penalty function~\eqref{eq:p_d} related to the AIC, MDL or HQC tests.

The overall procedure to find the sources number in the context of missing data is summarized in Algorithm~\ref{alg:structured_det}.
\begin{algorithm}
	\caption{Detection of number of sources in the context of missing-data and fixed rank constraint}
	\label{alg:structured_det}
	\textbf{Input:} $N,\; K,\; \bY,\; \bA_1, \dots, \bA_K$, $\btheta^{(0)}$, $\varepsilon_1$, $\varepsilon_2$, $K_1$, $p(d), \; d=1,...,K_1$.\\
	\textbf{Output:} A solution $\hat{d}_{EM}$ to Problem~(\ref{eq:criter_det_obs}).
	\begin{enumerate}[label={\theenumi:}]
		\item \textbf{for} $\tilde{d}=0,\dots, K_1$ \textbf{do}
		\begin{enumerate}[label=\alph*),leftmargin=2em]
			\item compute the estimate ${\hat{\btheta}_{\tilde{d}}}$ via Algorithm~\ref{alg:EM} using \eqref{eq:solution_str_detection_mstep} as solution to the M-step with $d=\tilde{d}$;
			\item compute $L_y({\hat{\btheta}_{\tilde{d}}})$ in~\eqref{eq:str_detection_decision_stat_Ly} using the estimate ${\hat{\btheta}_{\tilde{d}}}$.
		\end{enumerate}
		\textbf{end for}
		\item evaluate \[ \hat{d}_{EM} = \operatorname*{argmin}\limits_{\tilde{d} = 0, \dots, K_1}  L_y({\hat{\btheta}_{\tilde{d}}}) + p(\tilde{d}) ;\]
		\item output $\hat{d}_{EM}$.
	\end{enumerate}
\end{algorithm}

\subsubsection{{Rank constraint and centro-Hermitianity}}
Let us assume that the covariance matrix belongs to both the uncertainty sets (\ref{set:fixed_rank}) and (\ref{set:persymmetry}), i.e.
\begin{equation}\label{set:fixed_rank_and_persymm}
\mathcal{C} = \left\{\begin{matrix}
\bM = \sigma_n^2 \bI + \bV\bS_f \bV^\dagger \\ 
\bM = \bJ {\bM}^{*} \bJ \\ 
\bV\bS_f \bV^\dagger \succeq \bzero \\ 
\mbox{Rank}(\bV\bS_f \bV^\dagger) \le d\\
\sigma_n^2 > 0
\end{matrix}\right. ,
\end{equation}
where $\bV$, $\bS_f$, $d$, and $\sigma_n^2$ are defined as in~(\ref{set:fixed_rank}), whereas $\bJ$ is given by~(\ref{eq:J_exchange_matrix}).

Therefore, for a given $d$, the maximizer of $Q(\btheta, \btheta^{(h-1)})$ is given by~\cite{stoica-jansson-1999}
{\begin{equation}\label{eq:sol_det_M_step_pers_str}
		\begin{aligned}
         \btheta_{d,FB}^{(h)} = [&\tilde{\lambda}_{1, FB}, \dots, \tilde{\lambda}_{d, FB}, \tilde{\sigma}_{n, FB}^2,\\& \tilde{\bPhi}_{1,FB, R}, \tilde{\bPhi}_{1,FB, I}, \dots, \tilde{\bPhi}_{d,FB,R}, \tilde{\bPhi}_{d,FB,I} ]^{\mathrm{T}}, 
         \end{aligned}
\end{equation}}
where $\tilde{\lambda}_{v, FB}$ and $\tilde{\bPhi}_{v,FB}, \;v=1,\dots, d$ are the $d$ greatest eigenvalues and the corresponding eigenvectors of $\bSigma_{FB}$, defined as in~\eqref{eq:sigma_FB}, with $\tilde{\bPhi}_{v,FB,R}$ and $\tilde{\bPhi}_{v,FB,I}$ the real and imaginary components of $\tilde{\bPhi}_{v,FB}$, respectively, and 
\begin{equation}
    \tilde{\sigma}_{n, FB}^2 = \frac{1}{N-d} \sum_{v=N-d}^N \tilde{\lambda}_{v, FB}
\end{equation}
is the arithmetic mean of the $N - d$ smallest eigenvalues of $\bSigma_{FB}$.
As a consequence,
\begin{equation}\label{eq:estimate_det_centroh}
	\begin{aligned}
    \hat{\bM}(\hat{\btheta}_{d,FB}) =&  \sum_{v=1}^d (\tilde{\lambda}_{v, FB} - \tilde{\sigma}_{n, FB}^2) \tilde{\bPhi}_{v,FB} \tilde{\bPhi}_{v,FB}^\dagger +\\ & \quad \; + \tilde{\sigma}_{n, FB}^2 \; \bI
    \end{aligned}
\end{equation}
with $\hat{\btheta}_{d,FB}$ the resulting estimate of Algorithm~\ref{alg:EM}.

Along the same line as the previous case, the statistic is computed for each possible $d$, to get the order estimate
\begin{equation}\label{eq:det_str_pers_test}
    \hat{d}_{EM-FB} = \operatorname*{argmin}\limits_{d = 0, \dots, K_1}   L_y(\hat{\btheta}_{d,FB}) + \frac{1}{2} p(d) ,
\end{equation}
 where $L_y({\hat{\btheta}_{d,FB}})$ is given by~\eqref{eq:str_detection_decision_stat_Ly} evaluated in correspondence of the estimate~\eqref{eq:estimate_det_centroh} and $p(d)$ is one of the penalty functions in~\eqref{eq:p_d}~\cite{258125}.

\section{Performance Analysis}\label{section:perf_analysis}
\begin{figure*}[ht] 
    \centering
  \subfloat[\label{fig:rob_beam_sc_1_a}]{%
      \includegraphics[width=0.40\linewidth]{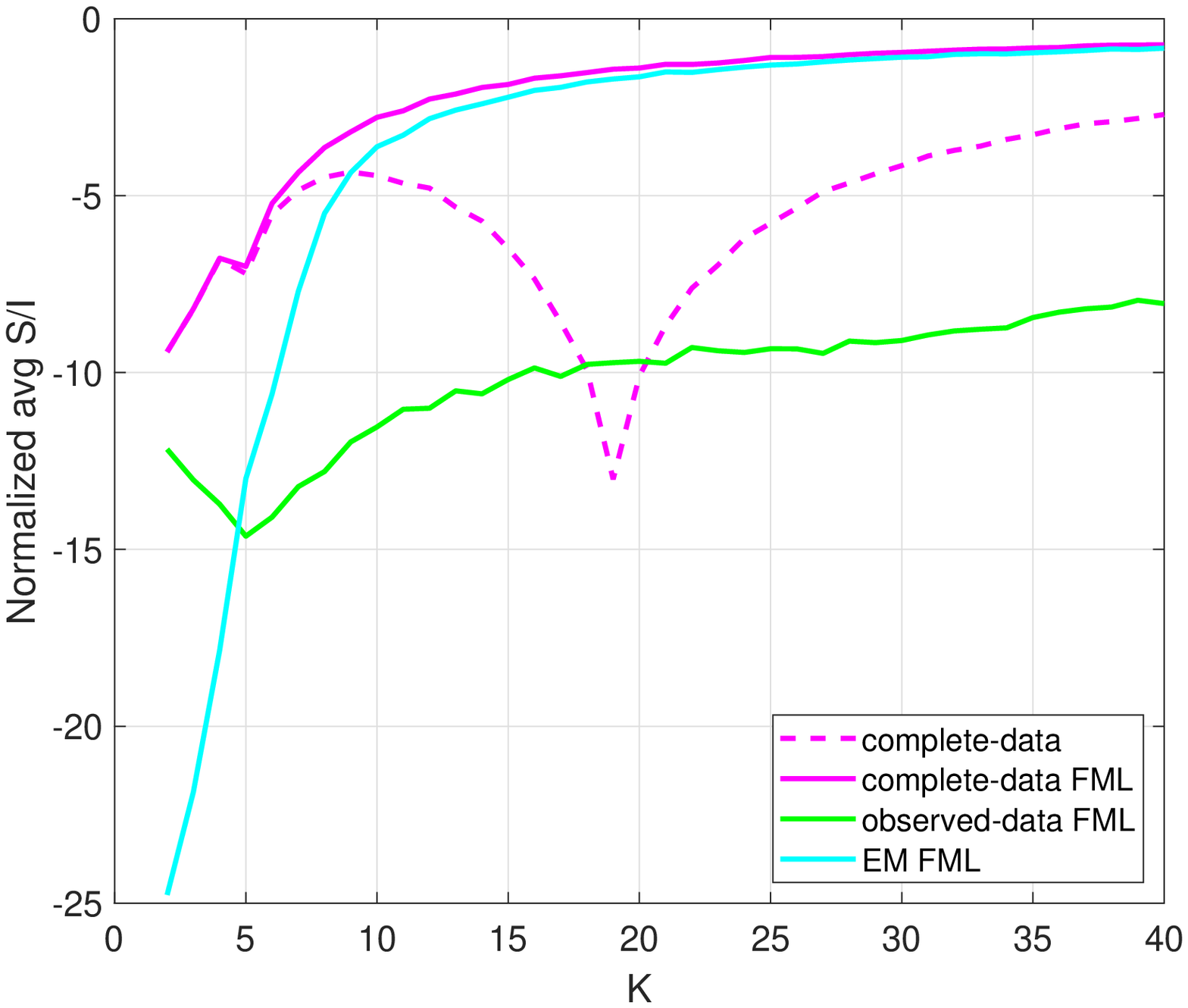}}
  \hspace{40pt}\subfloat[\label{fig:rob_beam_sc_1_b}]{%
        \includegraphics[width=0.40\linewidth]{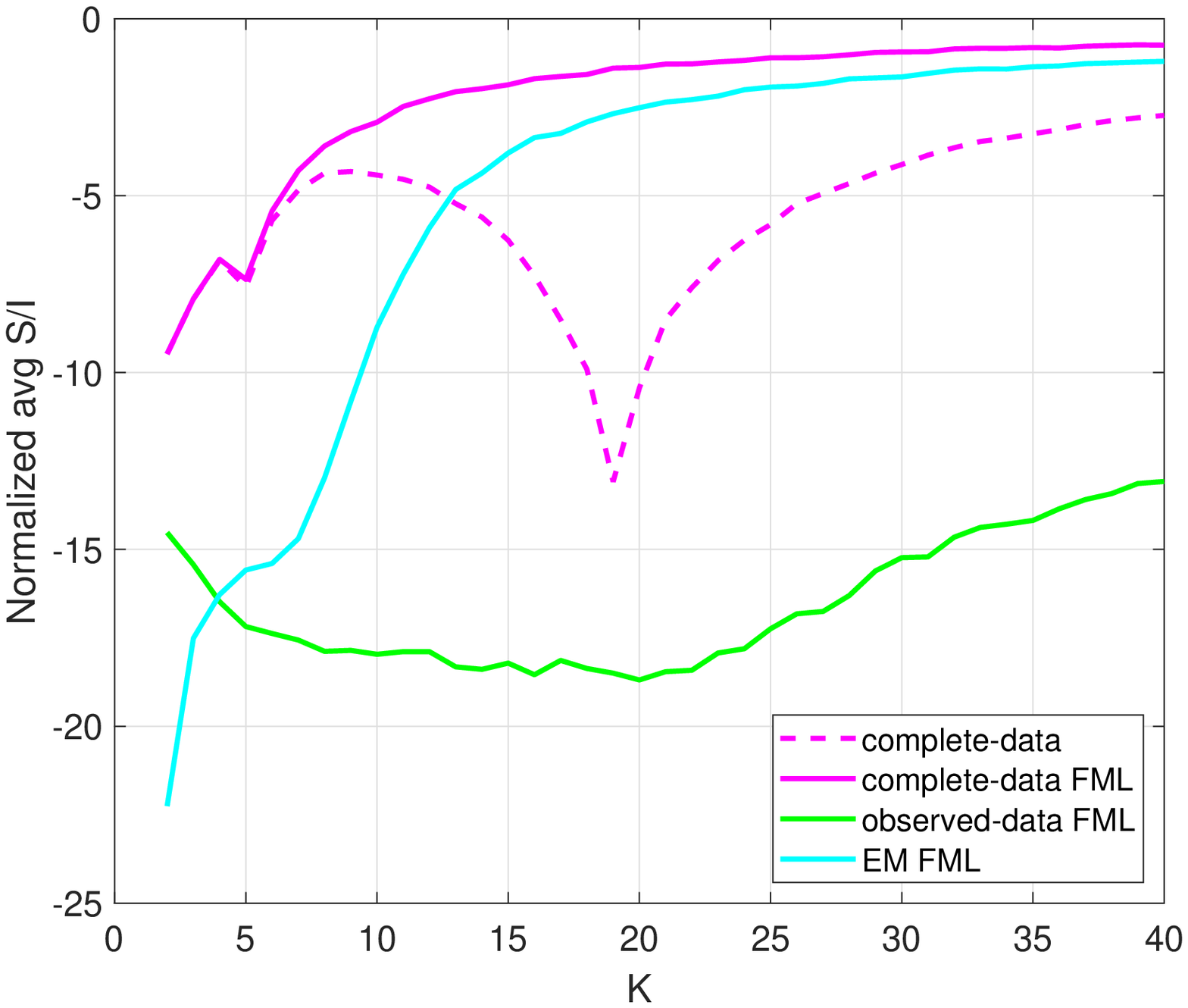}}
  \\
  \subfloat[\label{fig:rob_beam_sc_1_c}]{%
        \includegraphics[width=0.40\linewidth]{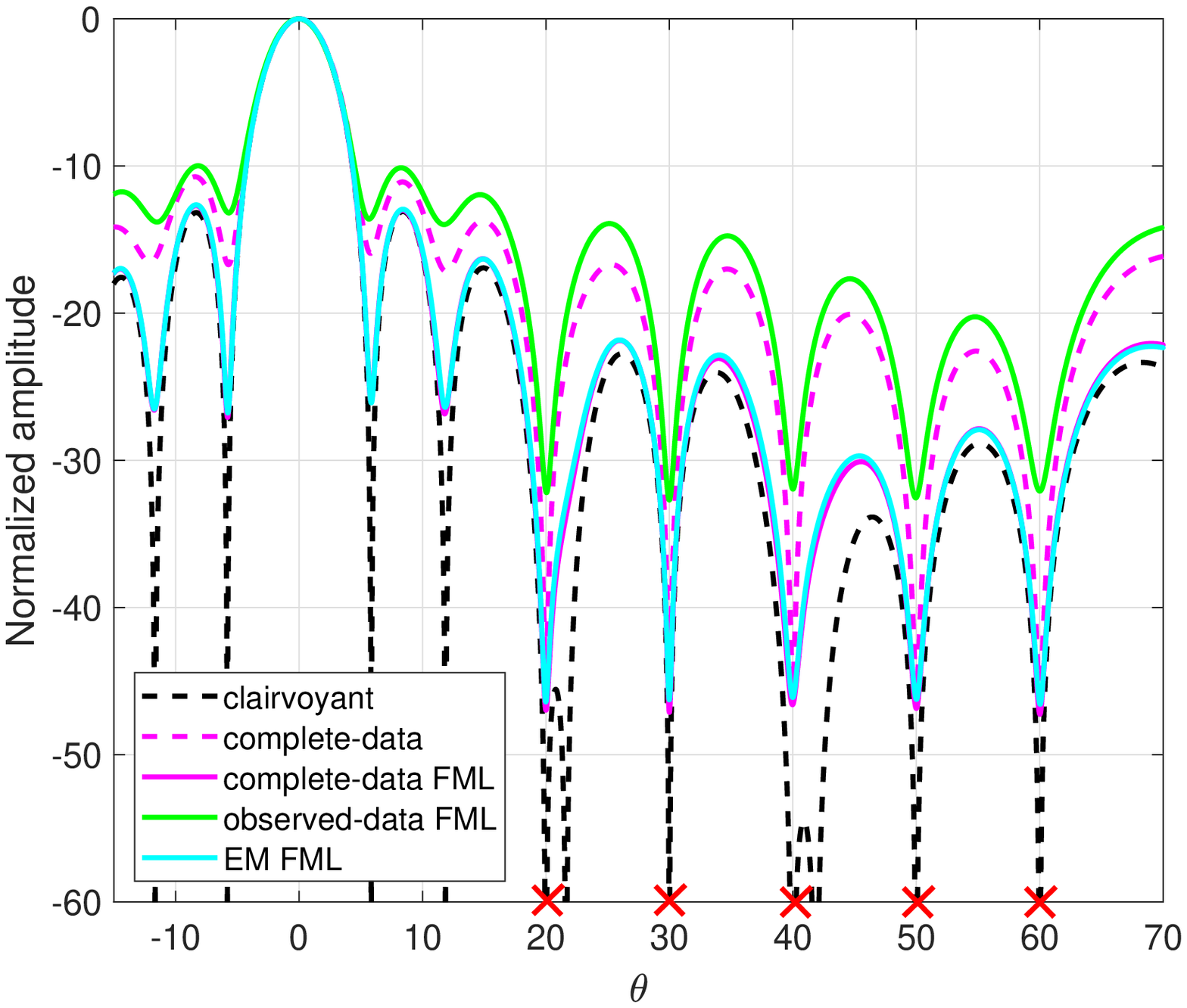}}
  \hspace{40pt}\subfloat[\label{fig:rob_beam_sc_1_d}]{%
        \includegraphics[width=0.40\linewidth]{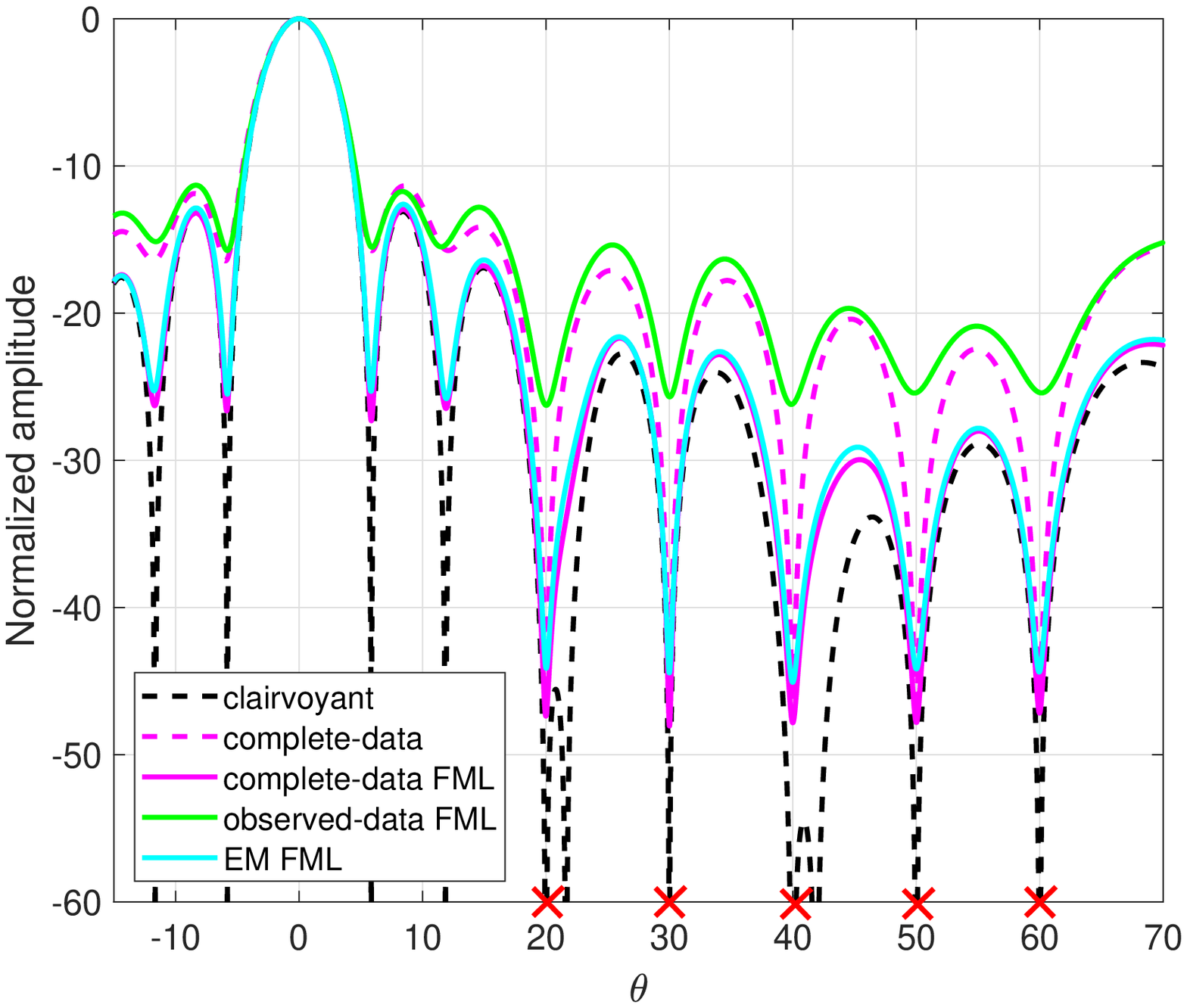}}
   \caption{Adaptive beamformer performance for a ULA with 20 antennas in Scenario 1. Figs. \subref{fig:rob_beam_sc_1_a} and \subref{fig:rob_beam_sc_1_c} consider $p_{\mathrm m} = 0.1$ while Figs. \subref{fig:rob_beam_sc_1_b} and \subref{fig:rob_beam_sc_1_d} consider $p_{\mathrm m} = 0.3$. Figs. \subref{fig:rob_beam_sc_1_a} and \subref{fig:rob_beam_sc_1_b} {display} the normalized average S/I versus number of snapshots, while Figs. \subref{fig:rob_beam_sc_1_c} and \subref{fig:rob_beam_sc_1_d} display the resulting beampattern with $K = 60$ (therein, the red-Xs along the $\theta$-axis denote the sources directions).}
  \label{fig:rob_beam_sc_1} 
\end{figure*}

\begin{figure*}[ht] 
    \centering
  \subfloat[\label{fig:rob_beam_sc_2_a}]{%
      \includegraphics[width=0.40\linewidth]{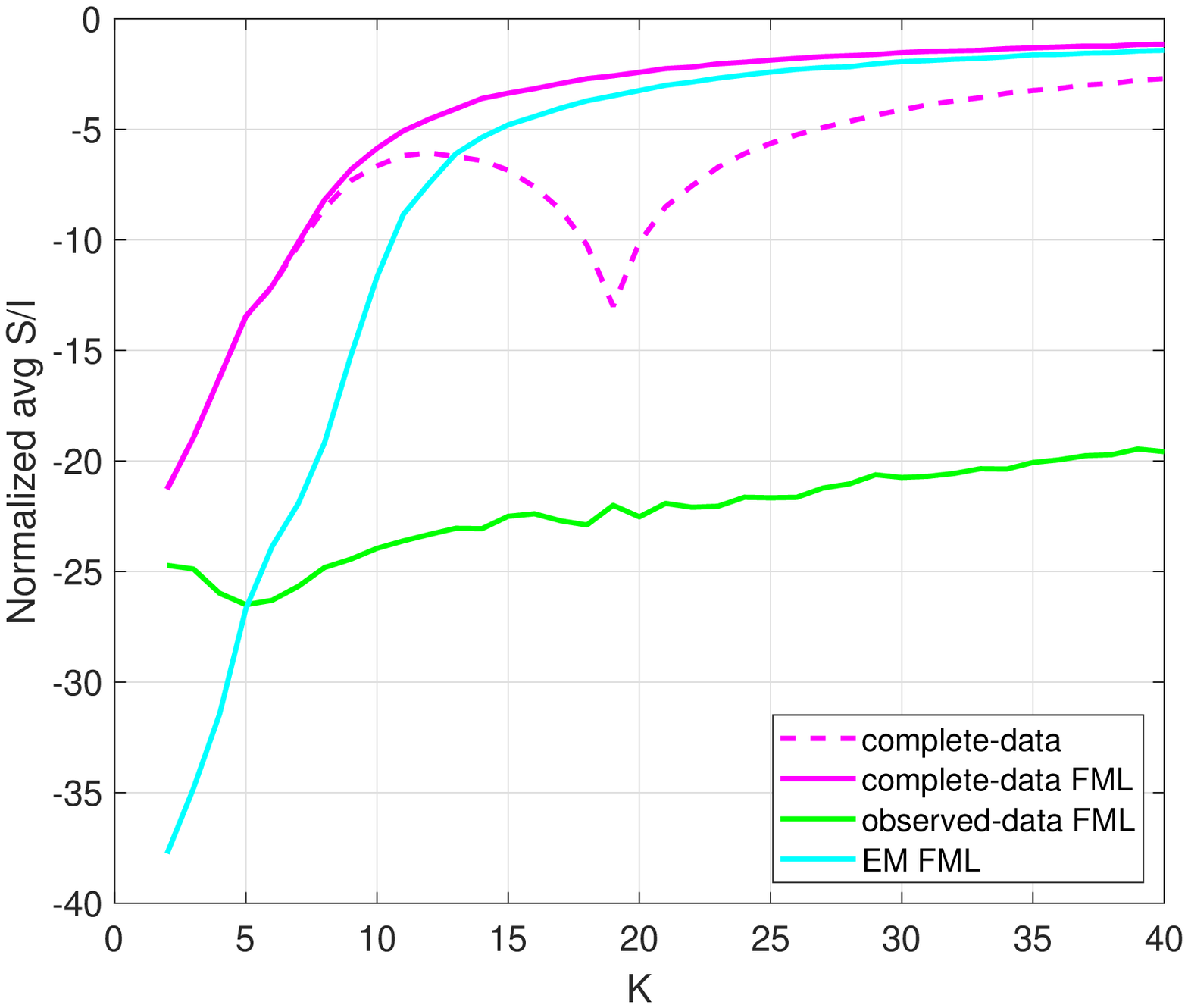}}
  \hspace{40pt}\subfloat[\label{fig:rob_beam_sc_2_b}]{%
        \includegraphics[width=0.40\linewidth]{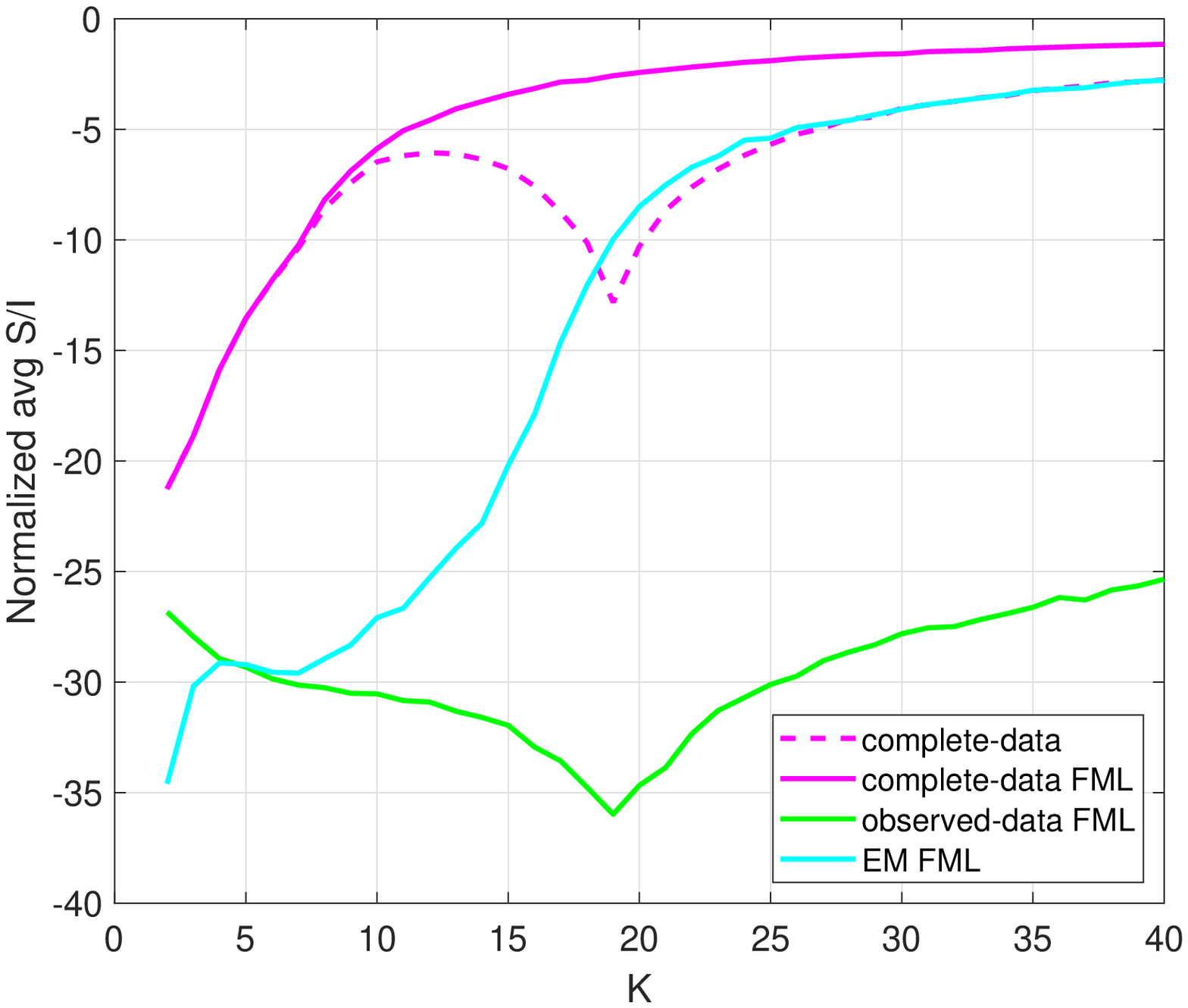}}
  \\
  \subfloat[\label{fig:rob_beam_sc_2_c}]{%
        \includegraphics[width=0.40\linewidth]{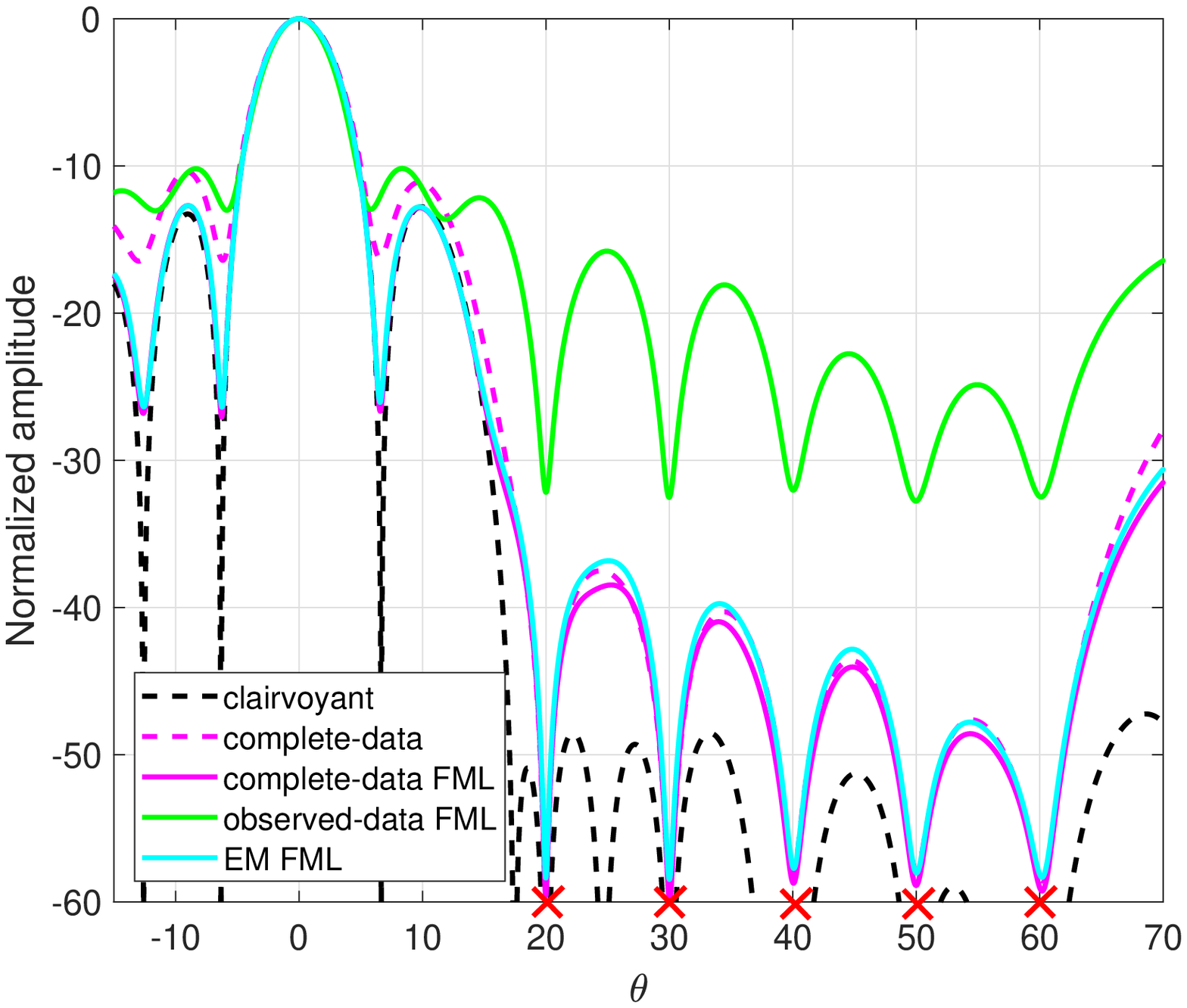}}
  \hspace{40pt}\subfloat[\label{fig:rob_beam_sc_2_d}]{%
        \includegraphics[width=0.40\linewidth]{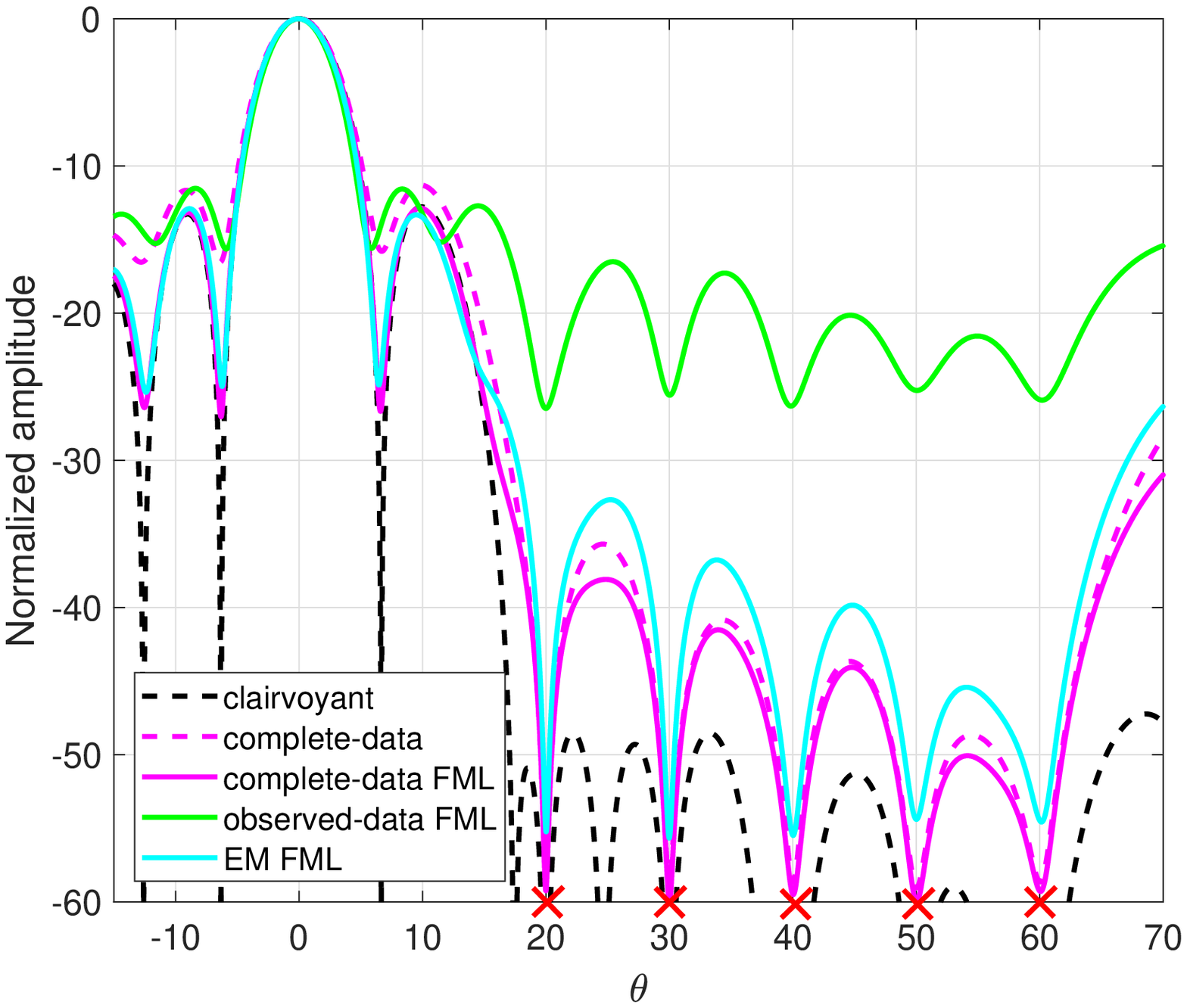}}
   \caption{Adaptive beamformer performance for a ULA with 20 antennas in Scenario 2. Figs. \subref{fig:rob_beam_sc_2_a} and \subref{fig:rob_beam_sc_2_c} consider $p_{\mathrm m} = 0.1$ while Figs. \subref{fig:rob_beam_sc_2_b} and \subref{fig:rob_beam_sc_2_d} consider $p_{\mathrm m} = 0.3$. Figs. \subref{fig:rob_beam_sc_2_a} and \subref{fig:rob_beam_sc_2_b} display the normalized average S/I versus number of snapshots, while Figs. \subref{fig:rob_beam_sc_2_c} and \subref{fig:rob_beam_sc_2_d} display the resulting beampattern with $K = 60$ (therein, the red-Xs along the $\theta$-axis denote the sources directions).}
  \label{fig:rob_beam_sc_2} 
\end{figure*}

\begin{figure*}[ht] 
    \centering
  \subfloat[\label{fig:sir_comparison_a}]{%
      \includegraphics[width=0.40\linewidth]{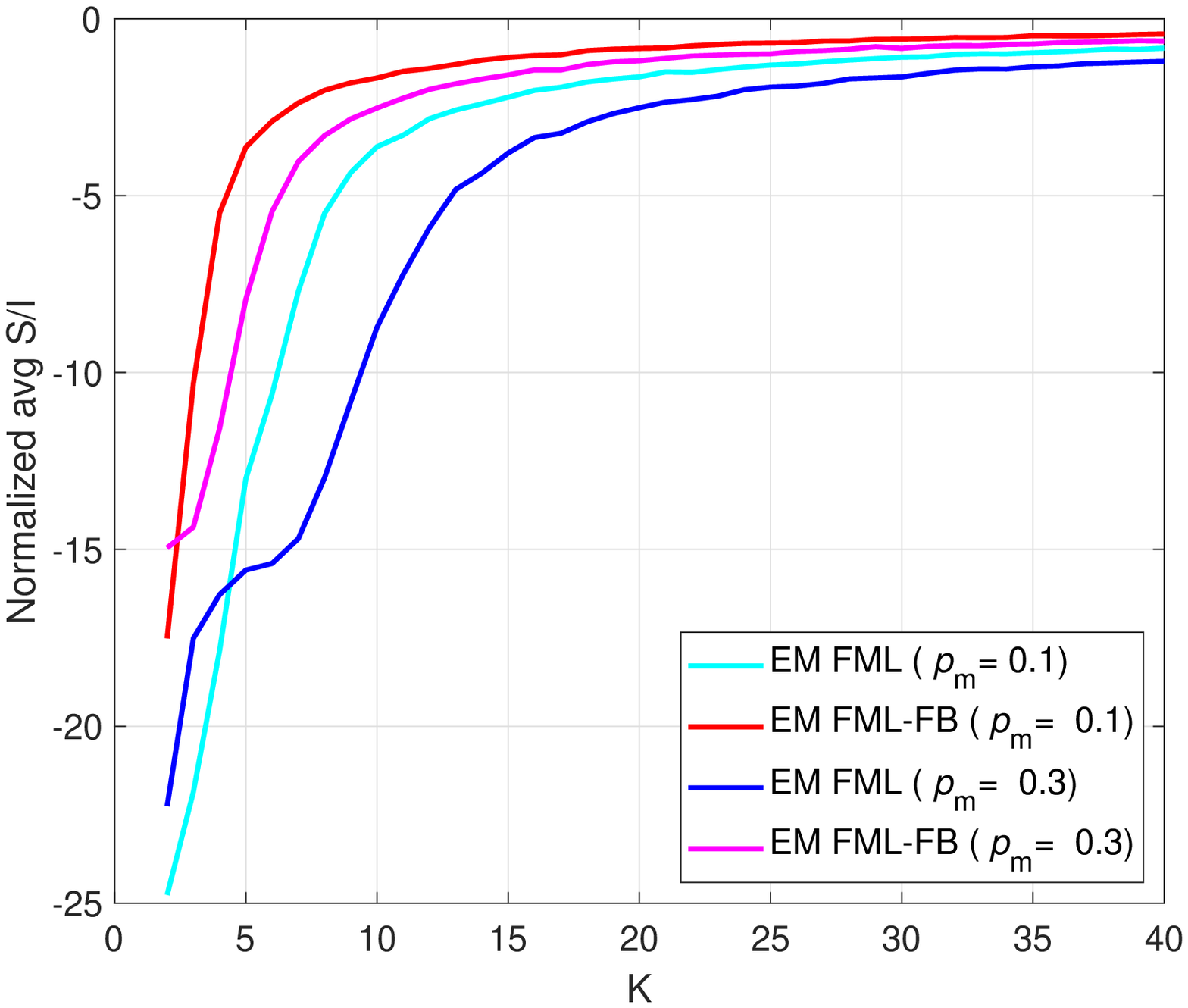}}
  \hspace{40pt}\subfloat[\label{fig:sir_comparison_b}]{%
        \includegraphics[width=0.40\linewidth]{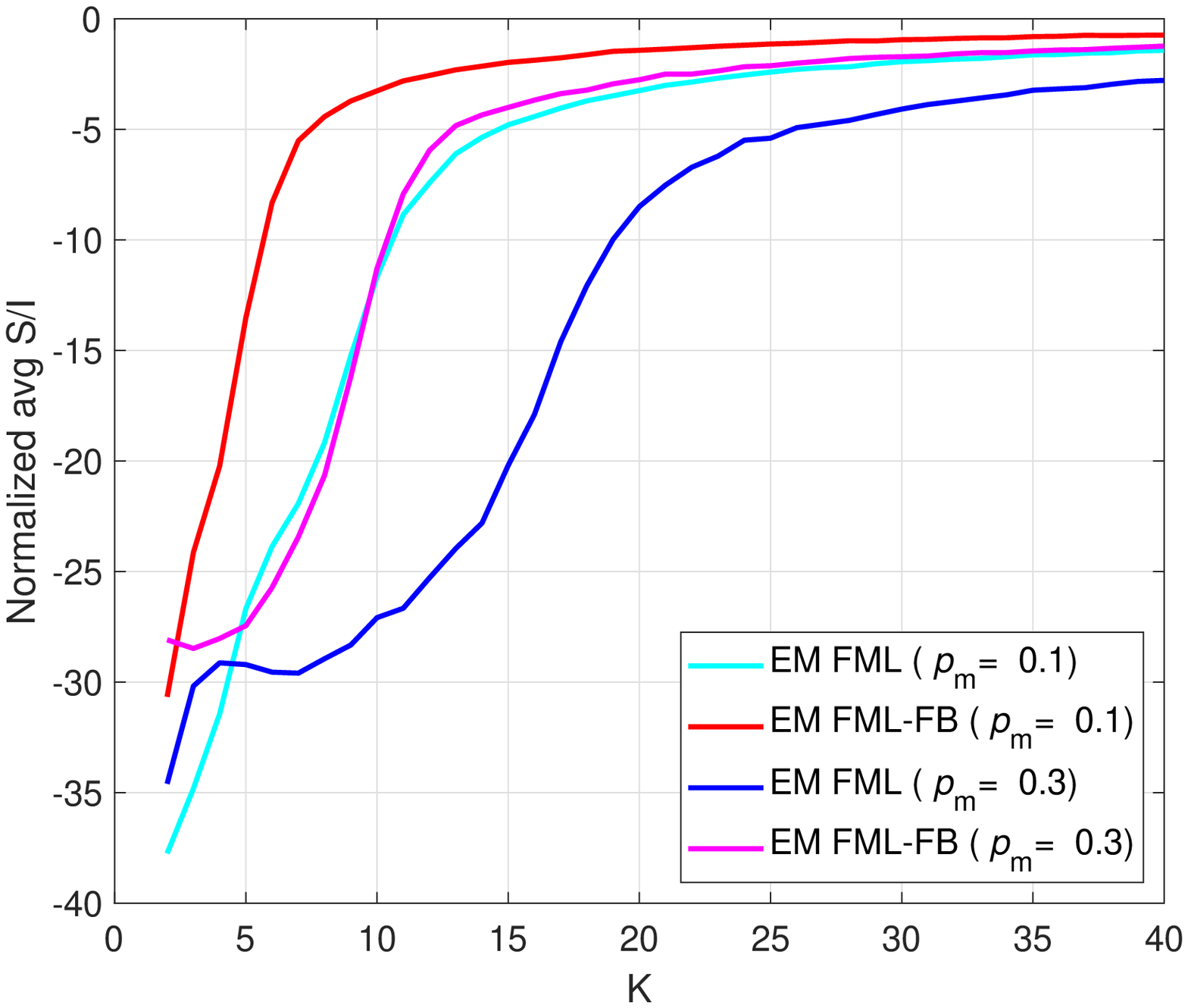}}
   \caption{Normalized average S/I versus number of snapshots for a ULA with 20 antennas. Fig.~\subref{fig:sir_comparison_a} considers Scenario 1 while Fig. \subref{fig:sir_comparison_b} Scenario 2.}
  \label{fig:sir_comparison} 
\end{figure*}

\begin{figure*}[ht] 
    \centering
  \subfloat[\label{fig:src_detection_pm_0_1_d2_AIC}]{%
\includegraphics[width=0.32\linewidth]{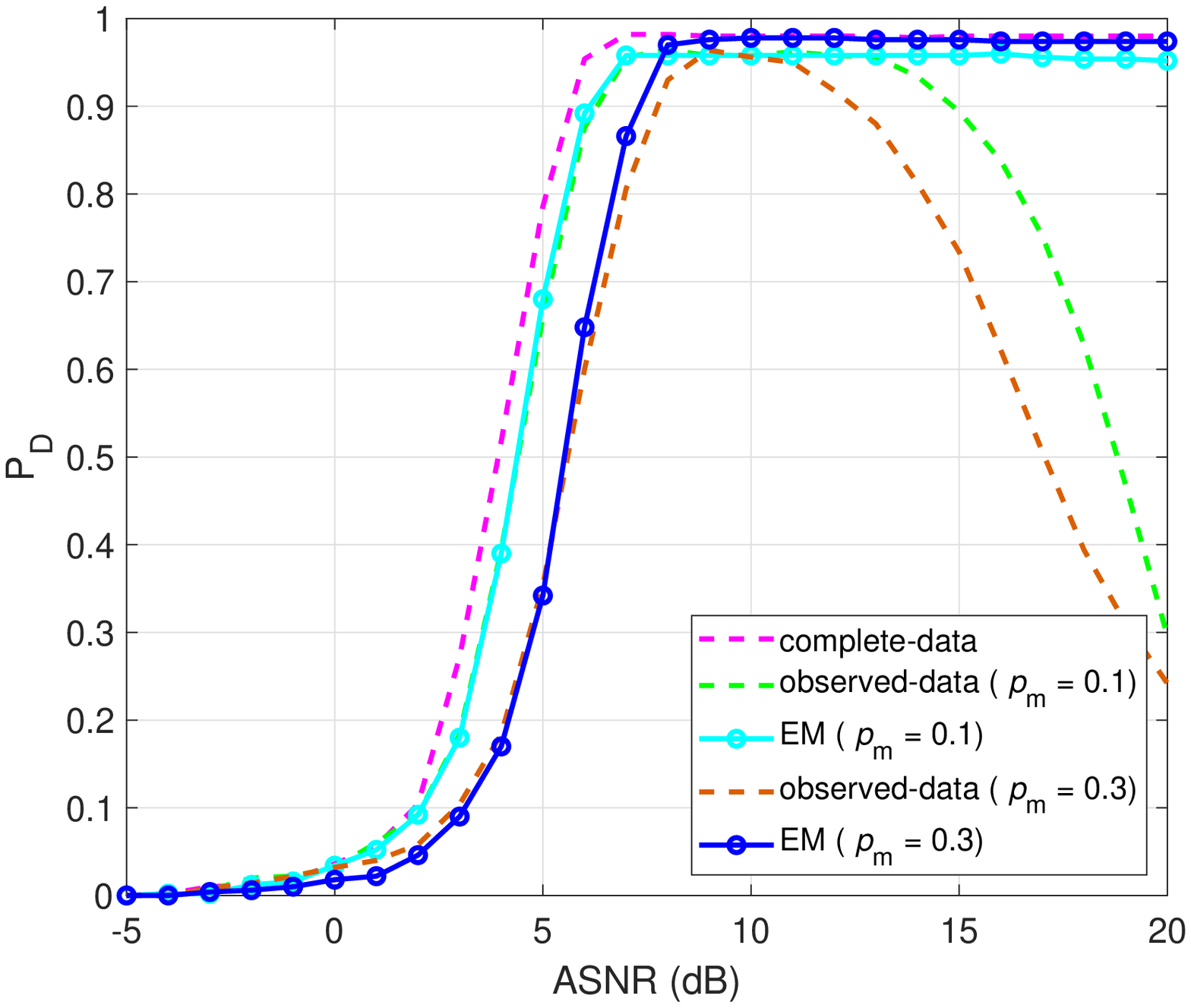}}
  \hspace{10pt}\subfloat[\label{fig:src_detection_pm_0_1_d2_MDL}]{%
      \includegraphics[width=0.32\linewidth]{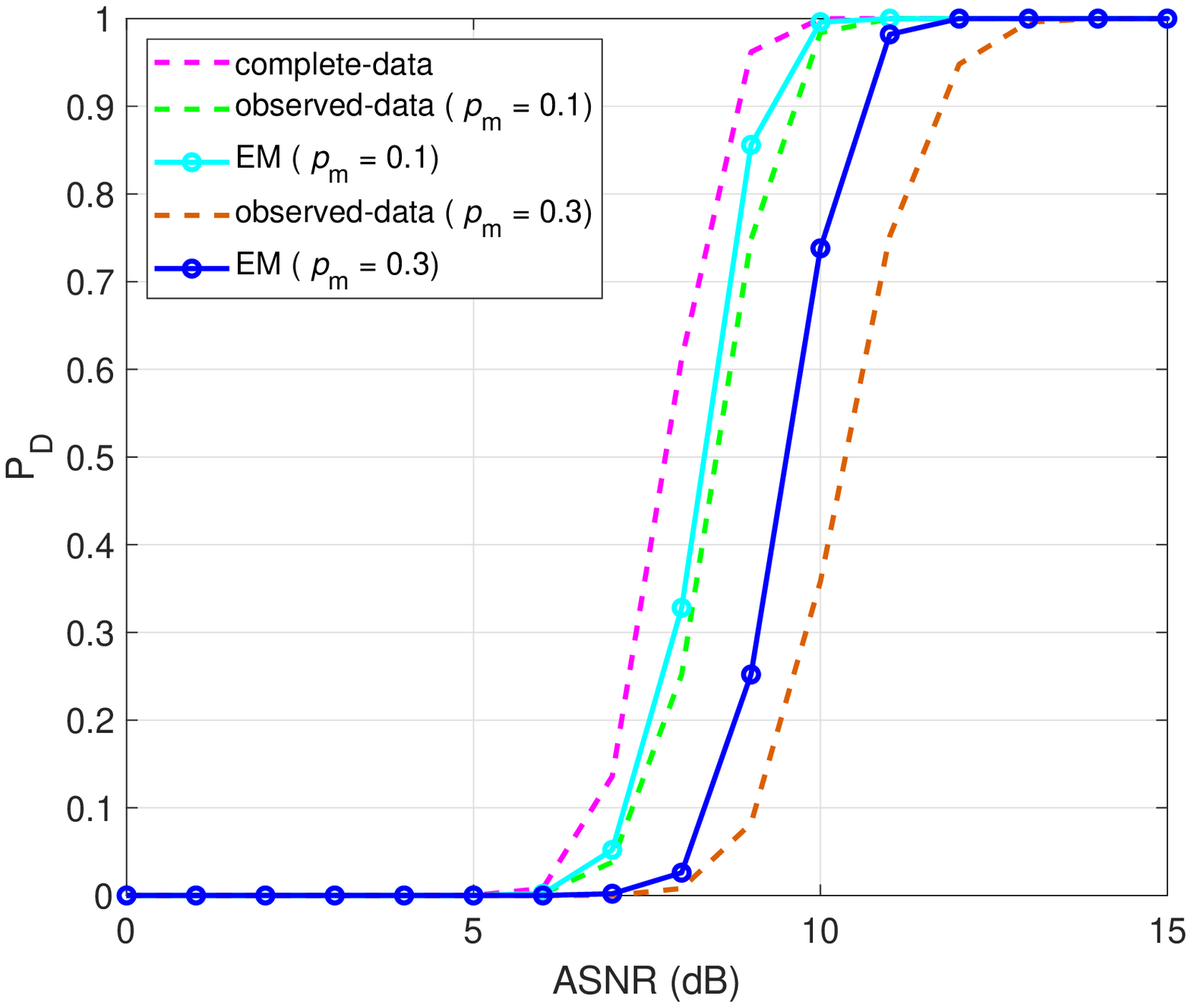}}
       \hspace{10pt}\subfloat[\label{fig:src_detection_pm_0_1_d2_HQC}]{%
        \includegraphics[width=0.32\linewidth]{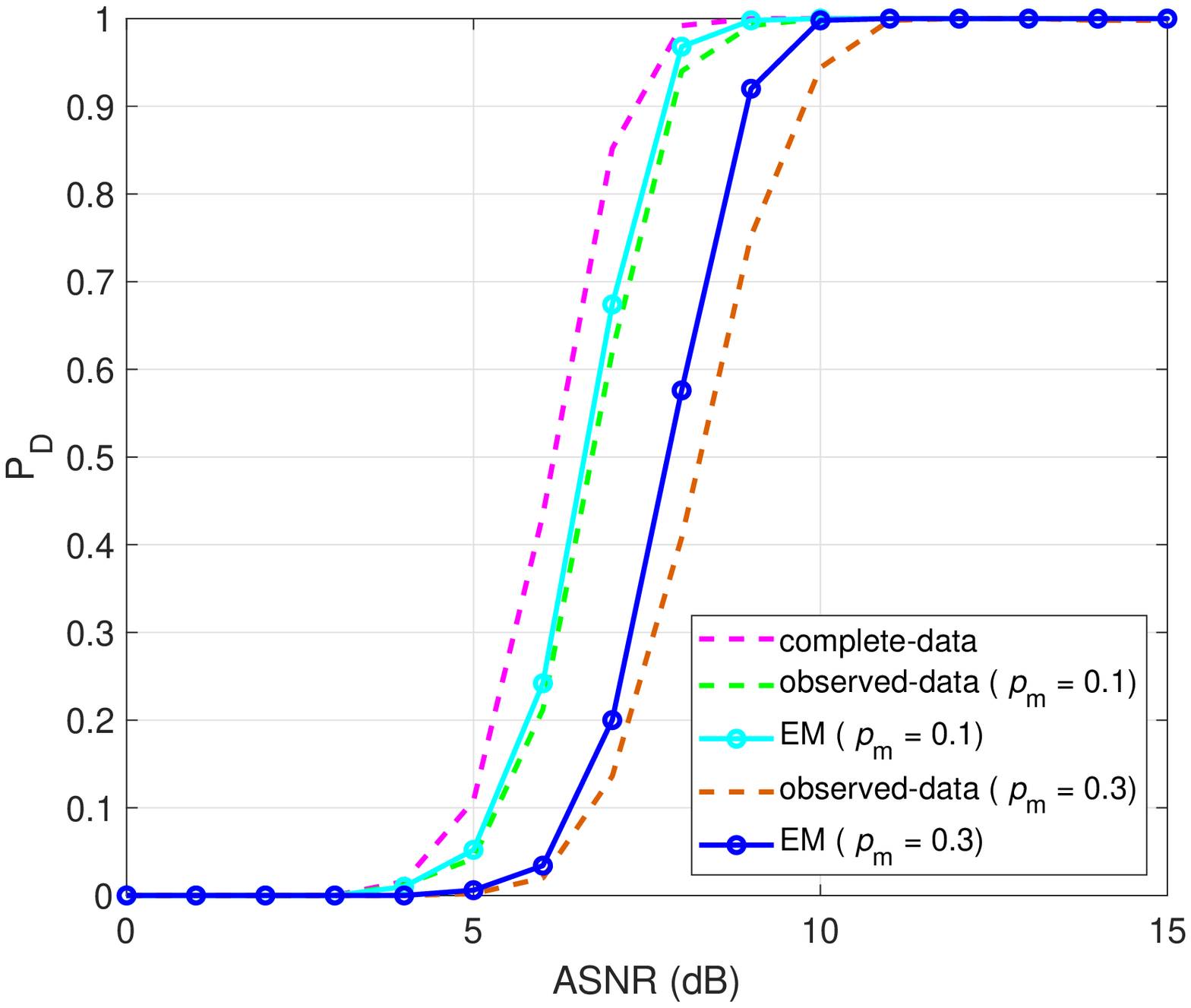}}\\
       \subfloat[\label{fig:src_detection_pm_0_1_d3_AIC}]{%
      \includegraphics[width=0.32\linewidth]{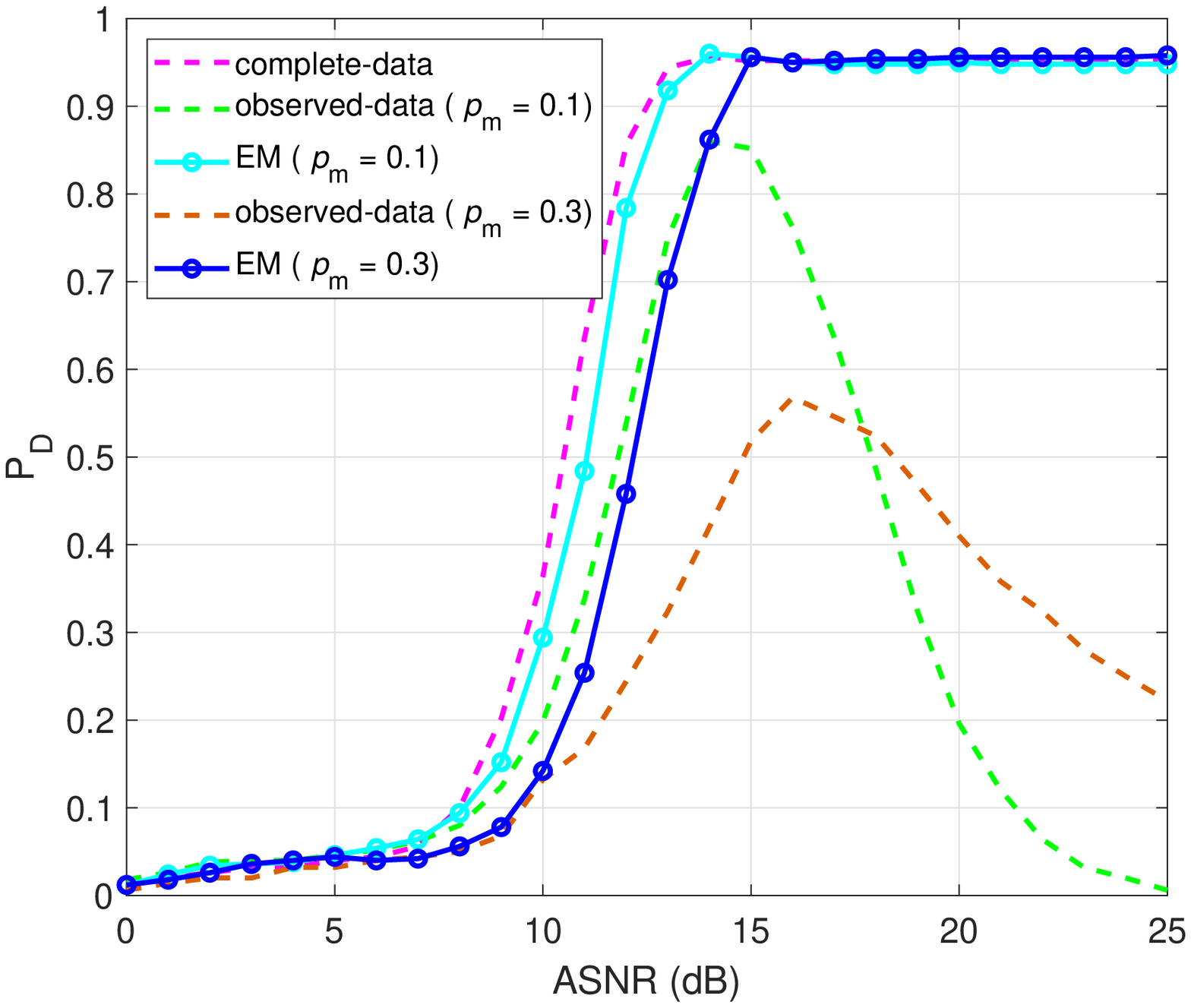}}
  \hspace{10pt}\subfloat[\label{fig:src_detection_pm_0_1_d3_MDL}]{%
      \includegraphics[width=0.32\linewidth]{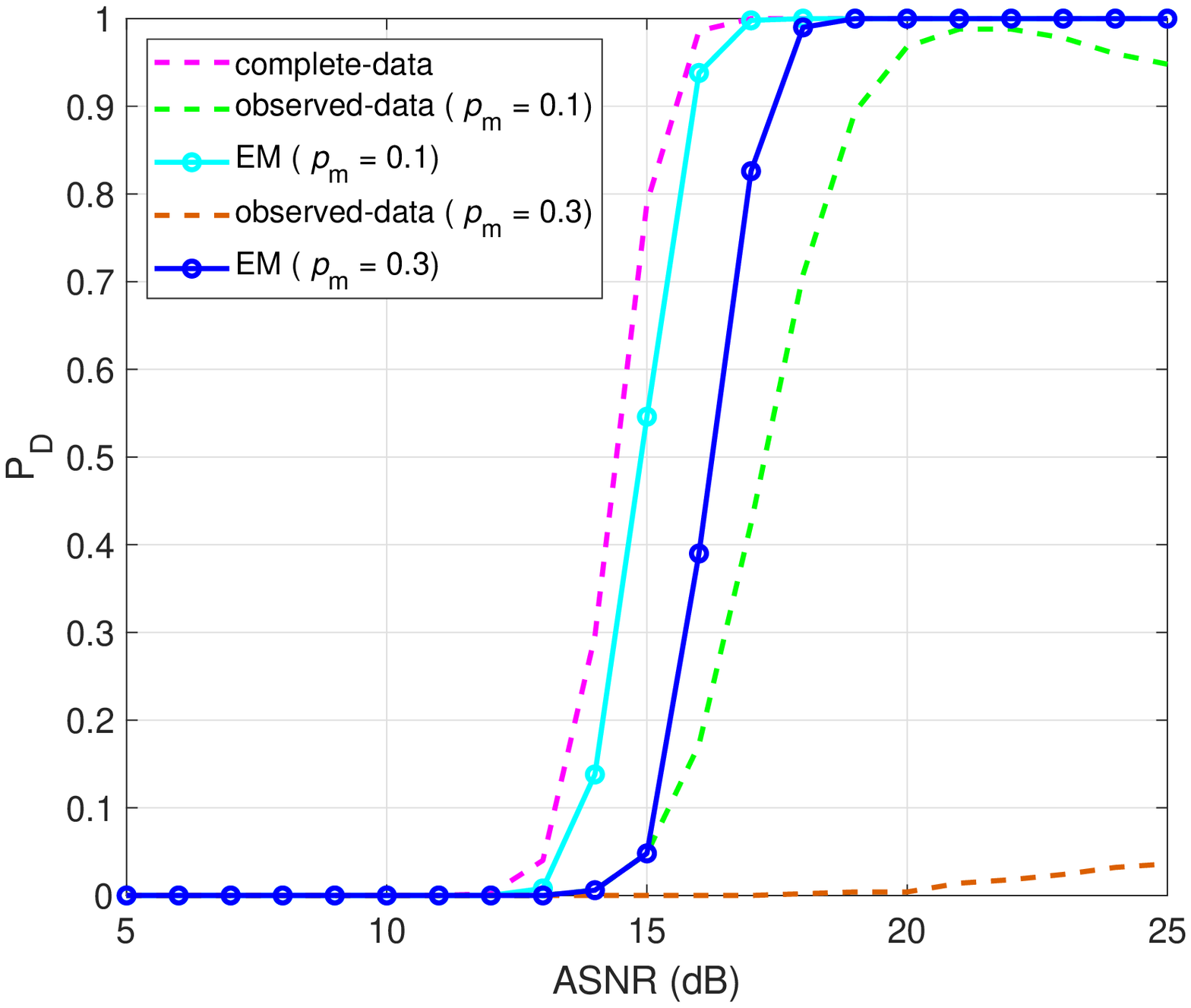}}
       \hspace{10pt}\subfloat[\label{fig:src_detection_pm_0_1_d3_HQC}]{%
        \includegraphics[width=0.32\linewidth]{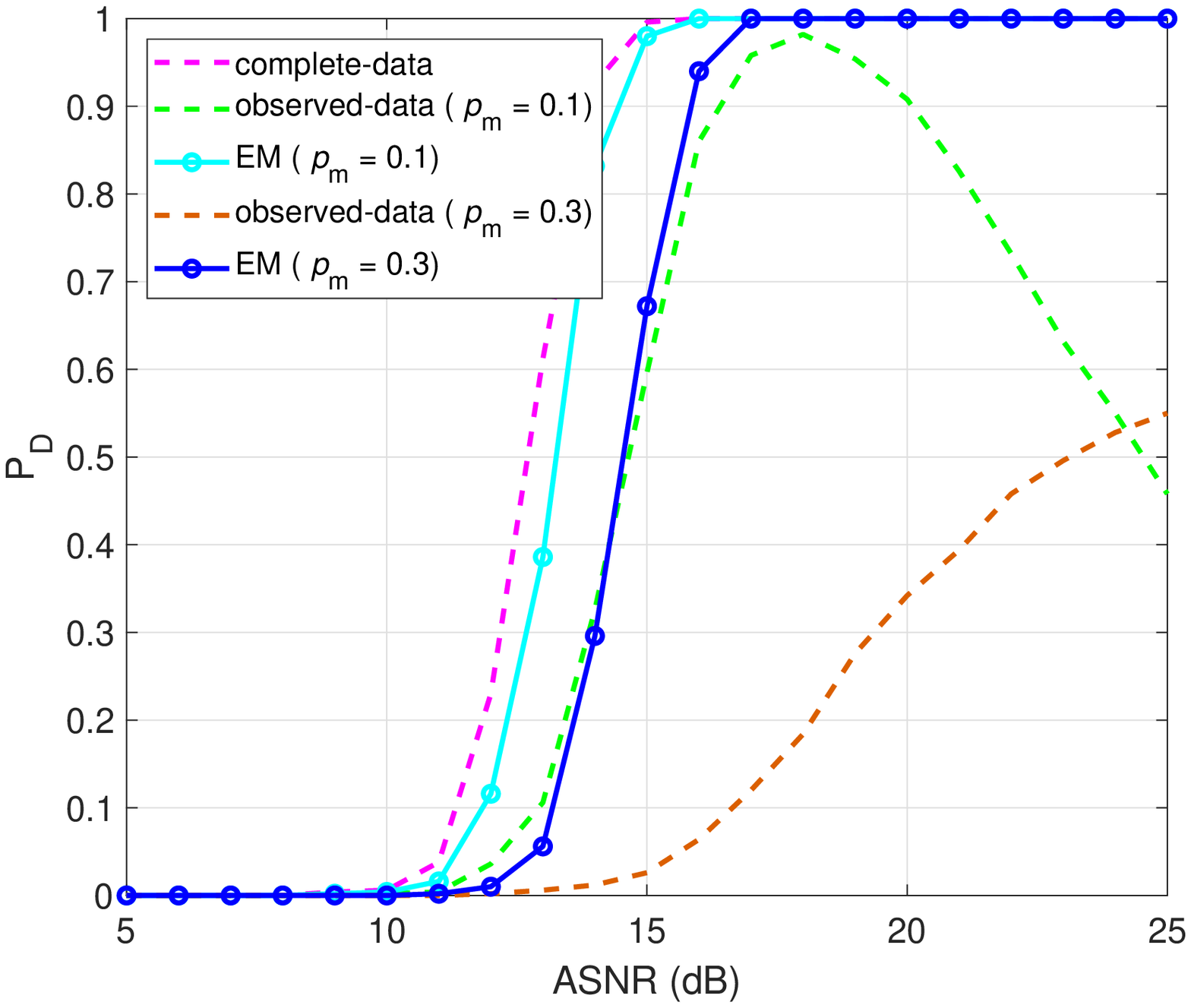}}\\
        \subfloat[\label{fig:src_detection_pm_0_1_d4_AIC}]{%
      \includegraphics[width=0.32\linewidth]{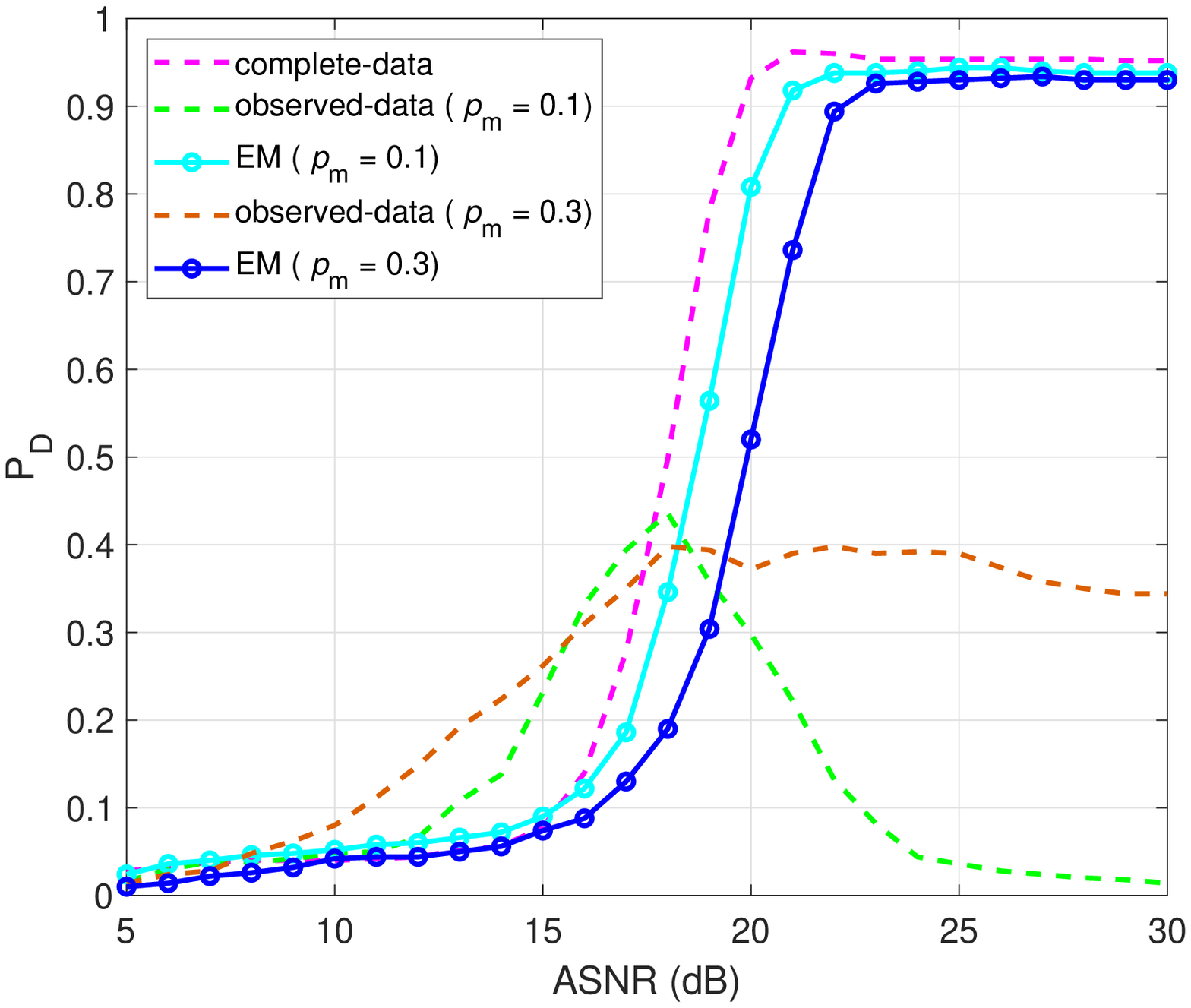}}
  \hspace{10pt}\subfloat[\label{fig:src_detection_pm_0_1_d4_MDL}]{%
      \includegraphics[width=0.32\linewidth]{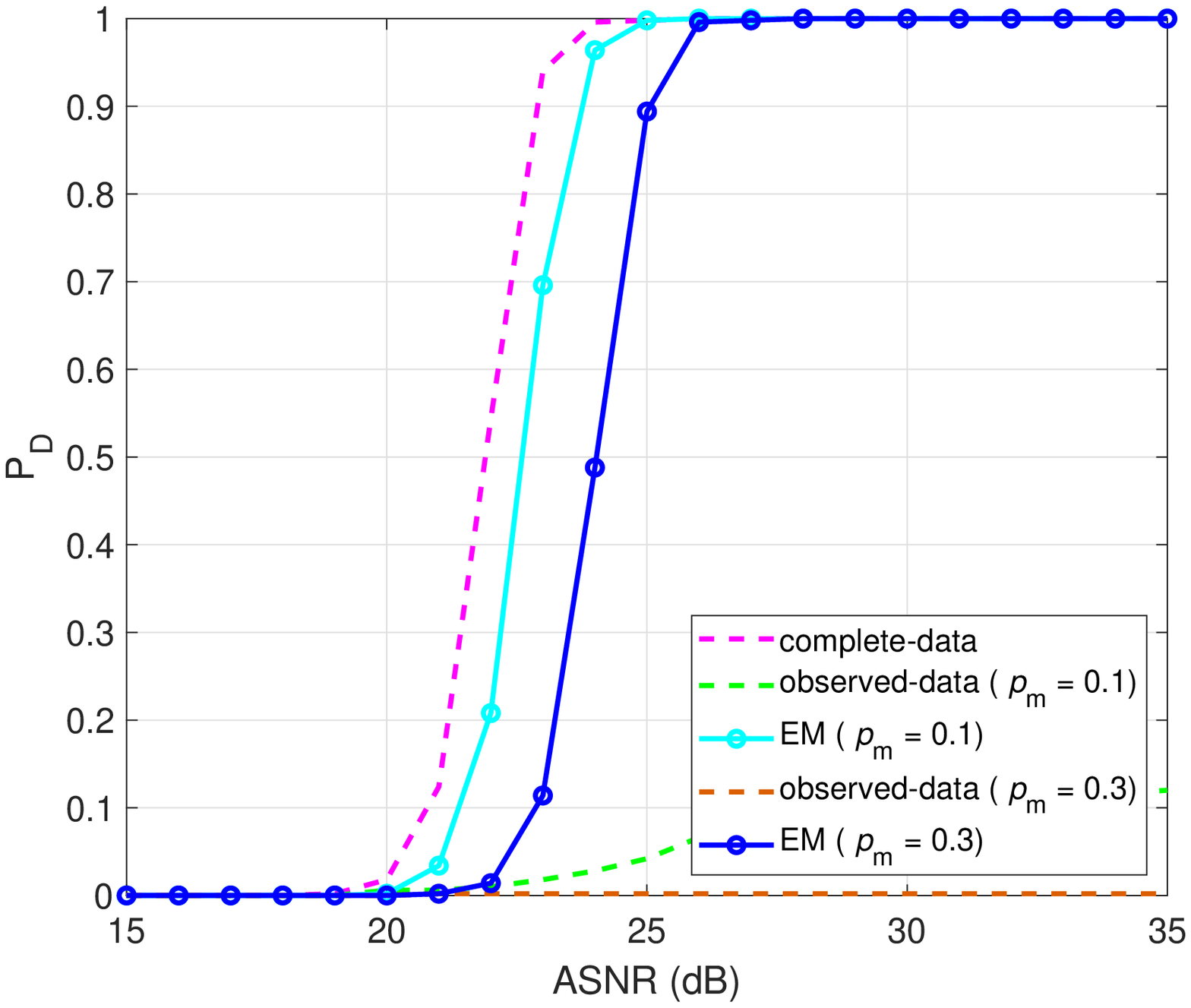}}
       \hspace{10pt}\subfloat[\label{fig:src_detection_pm_0_1_d4_HQC}]{%
        \includegraphics[width=0.32\linewidth]{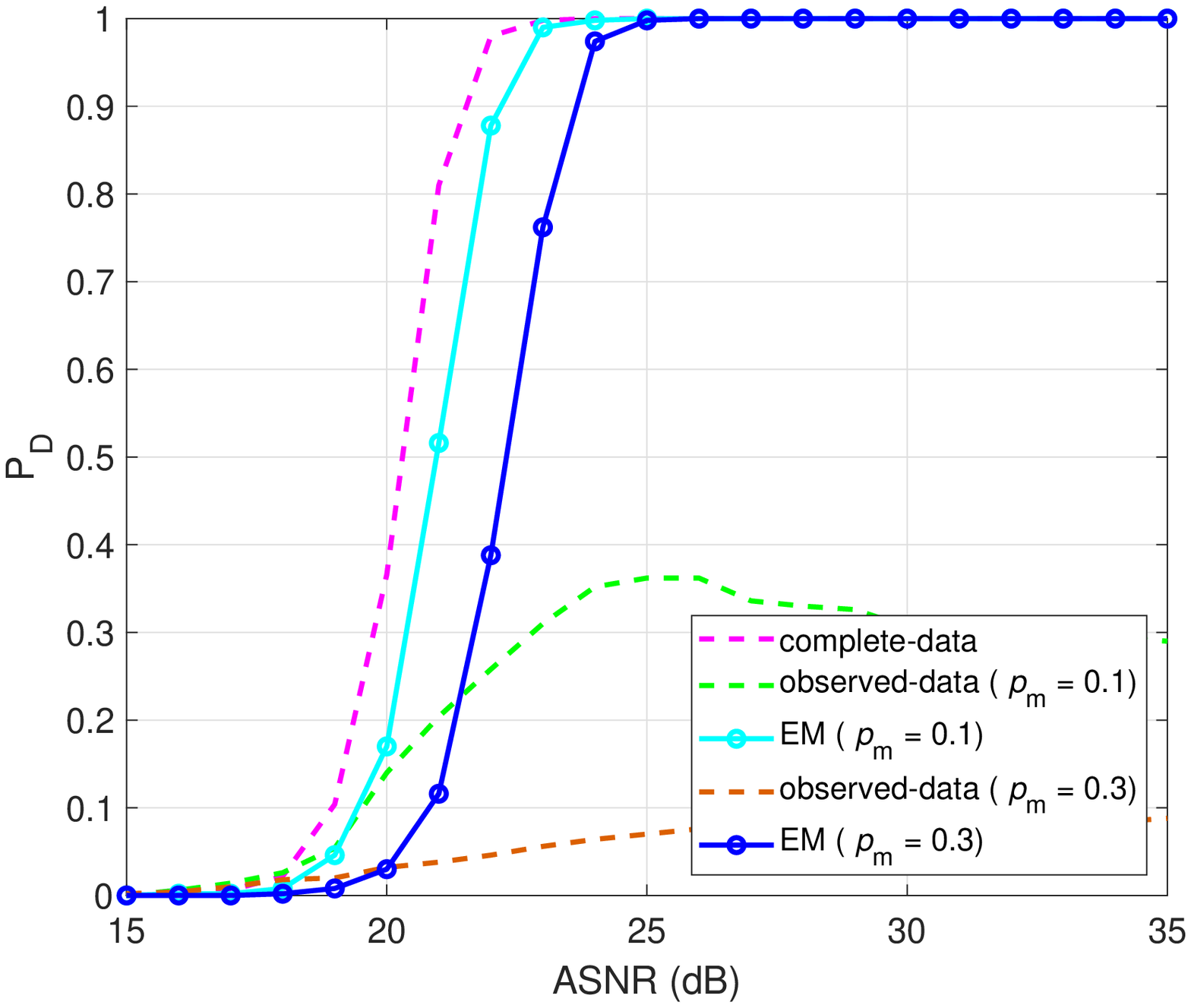}}
   \caption{Detection performance for a ULA with 20 antennas assuming $K=100$ and $p_{\mathrm m} \in \{0.1, 0.3\}$. Figs. (a), (b), and (c) assume $d=2$, Figs. (d), (e), and (f) assume $d=3$, whereas Figs. (g), (h), and (i) assume $d=4$ equal-power signals impinging the array, respectively, with signal separation corresponding to $0.891/N$. 
   Moreover, Figs. (a), (d), and (g) consider AIC, Figs. (b), (e), and (h) consider MDL, whereas Figs. (c), (f), and (i) consider HQC.}
  \label{fig:src_detection_1} 
\end{figure*}

\begin{figure*}[ht] 
   \centering
  \subfloat[\label{fig:src_detection_comp_pm_0_1_d3_AIC}]{%
\includegraphics[width=0.32\linewidth]{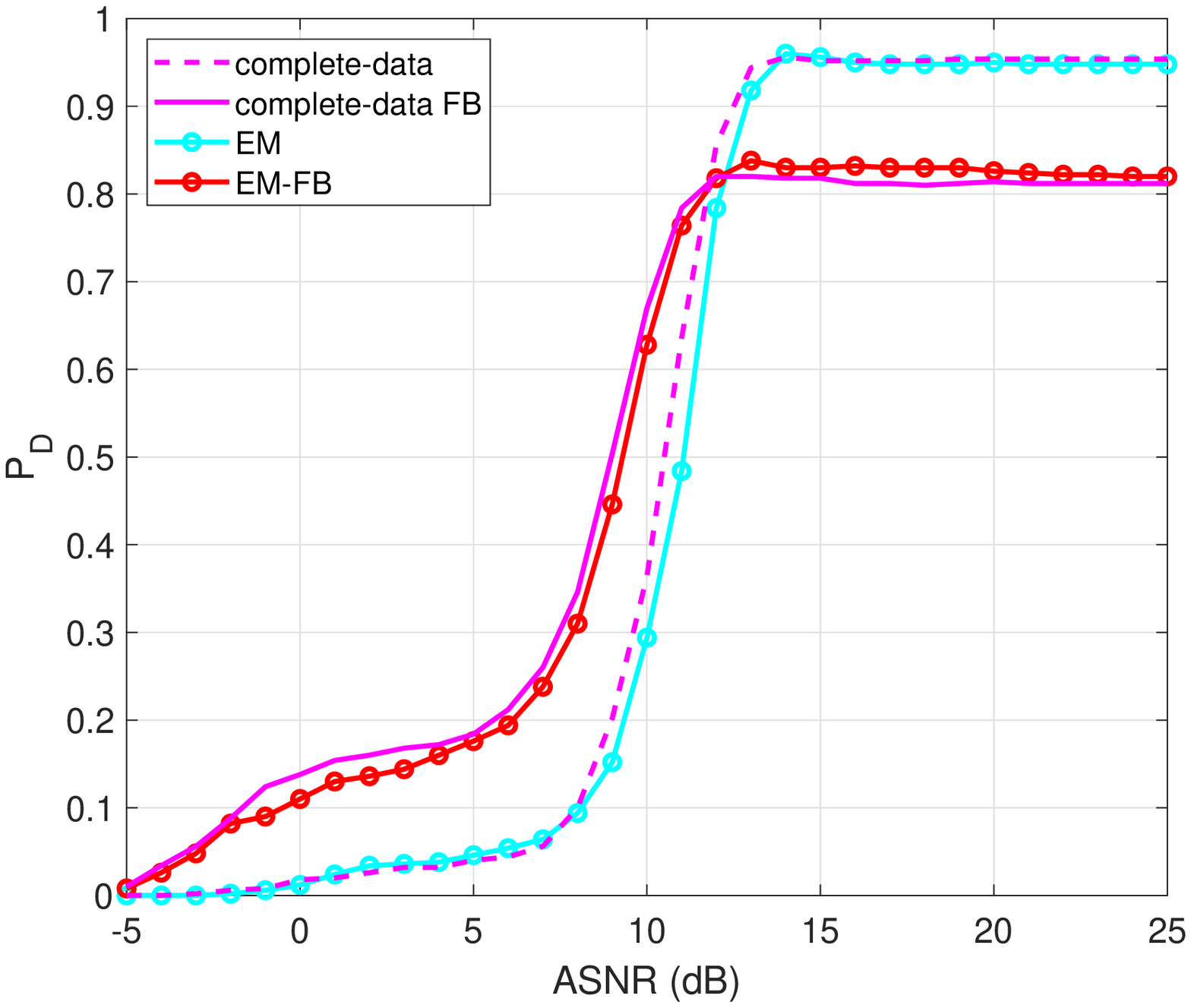}}
  \hspace{10pt}\subfloat[\label{fig:src_detection_comp_pm_0_1_d3_MDL}]{%
      \includegraphics[width=0.32\linewidth]{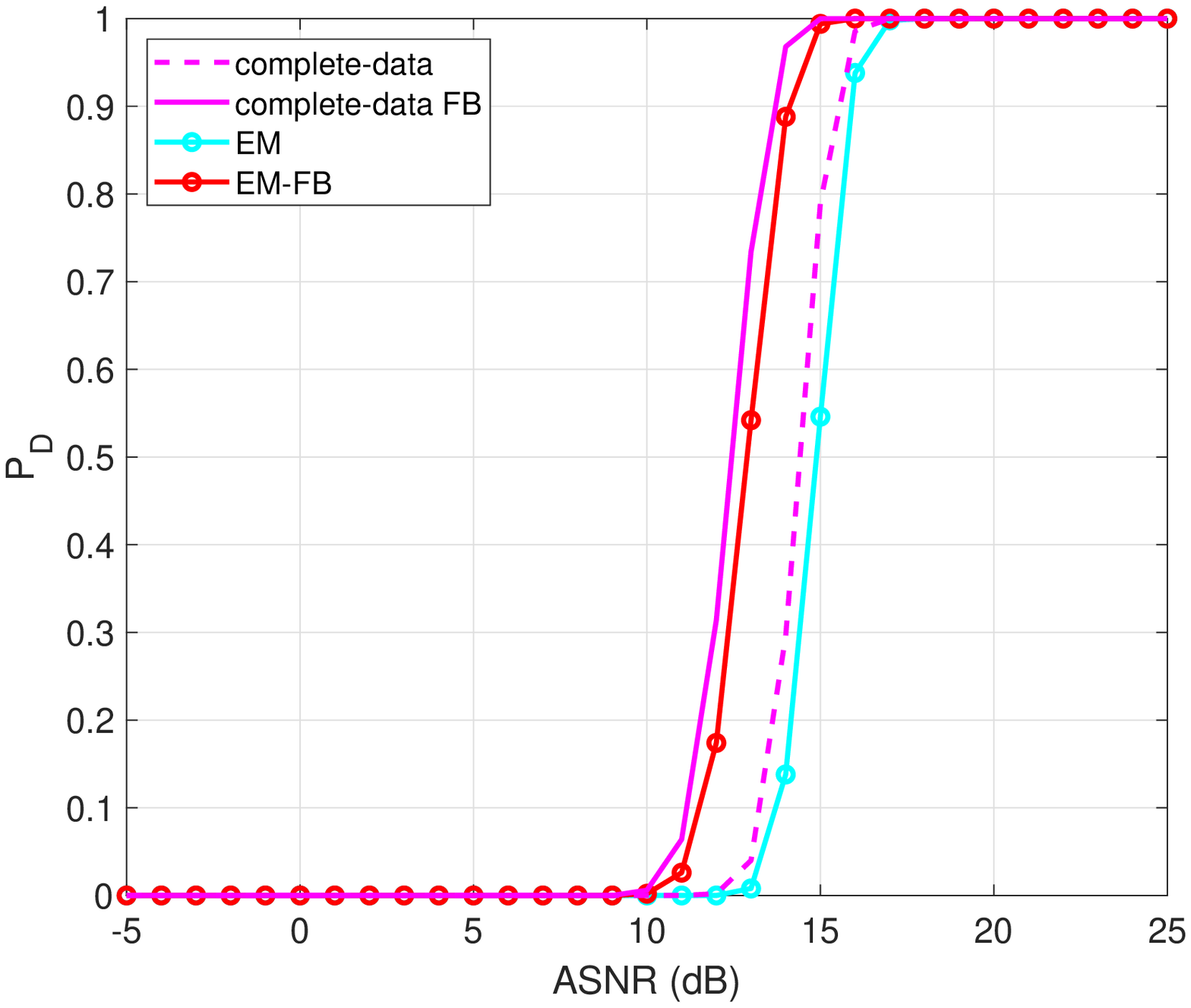}}
       \hspace{10pt}\subfloat[\label{fig:src_detection_comp_pm_0_1_d3_HQC}]{%
        \includegraphics[width=0.32\linewidth]{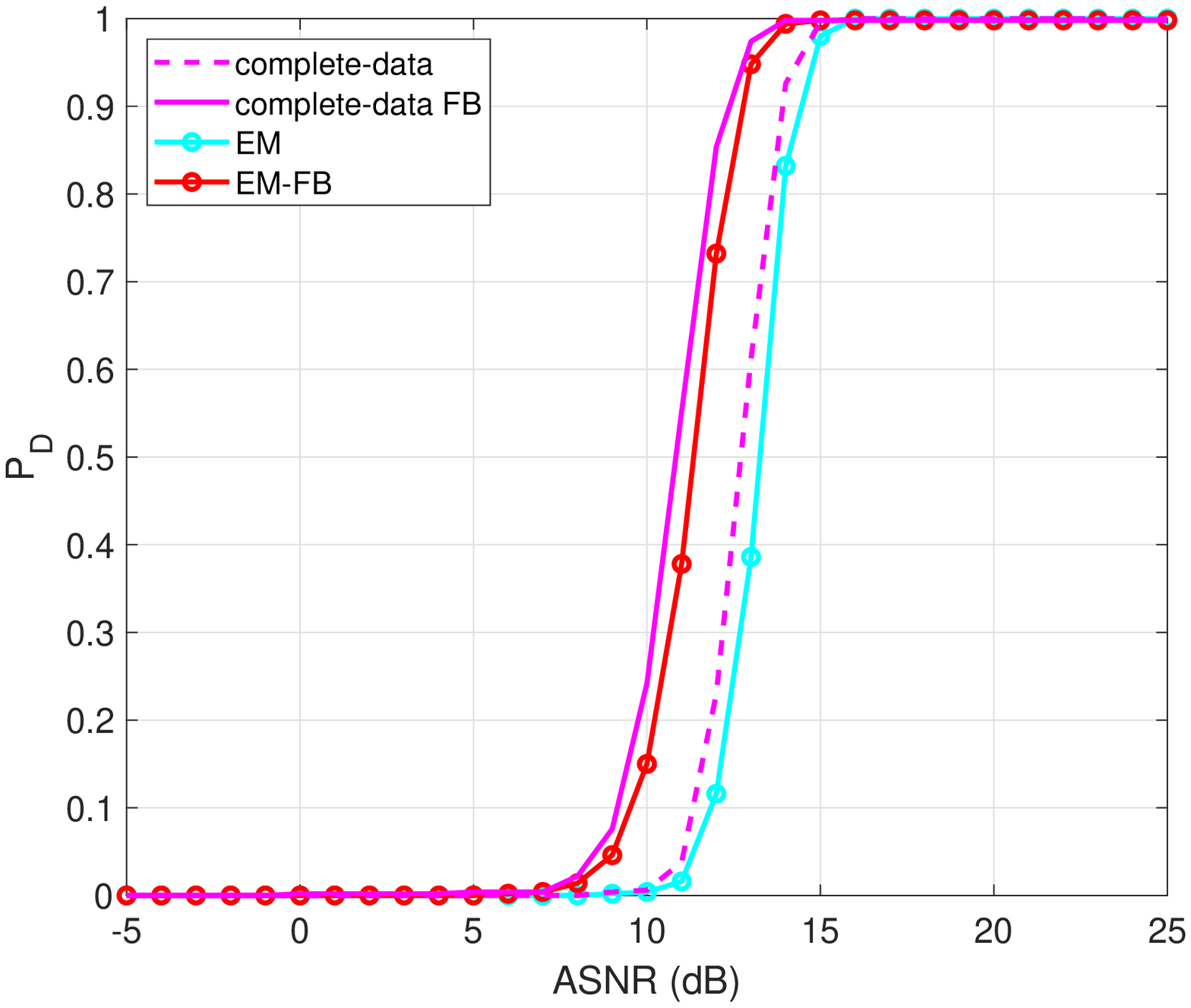}}\\
       \subfloat[\label{fig:src_detection_comp_pm_0_3_d3_AIC}]{%
      \includegraphics[width=0.32\linewidth]{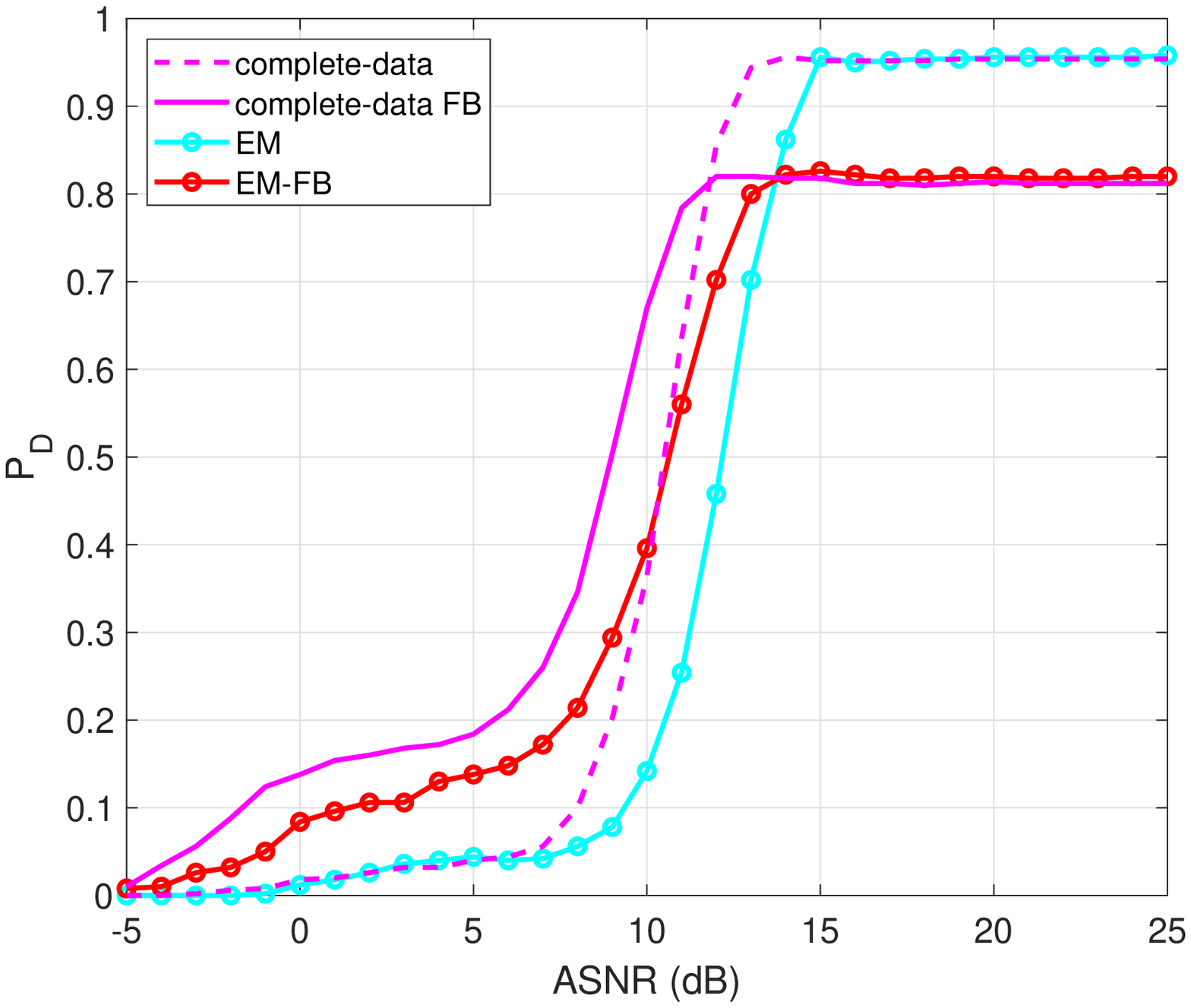}}
  \hspace{10pt}\subfloat[\label{fig:src_detection_comp_pm_0_3_d3_MDL}]{%
      \includegraphics[width=0.32\linewidth]{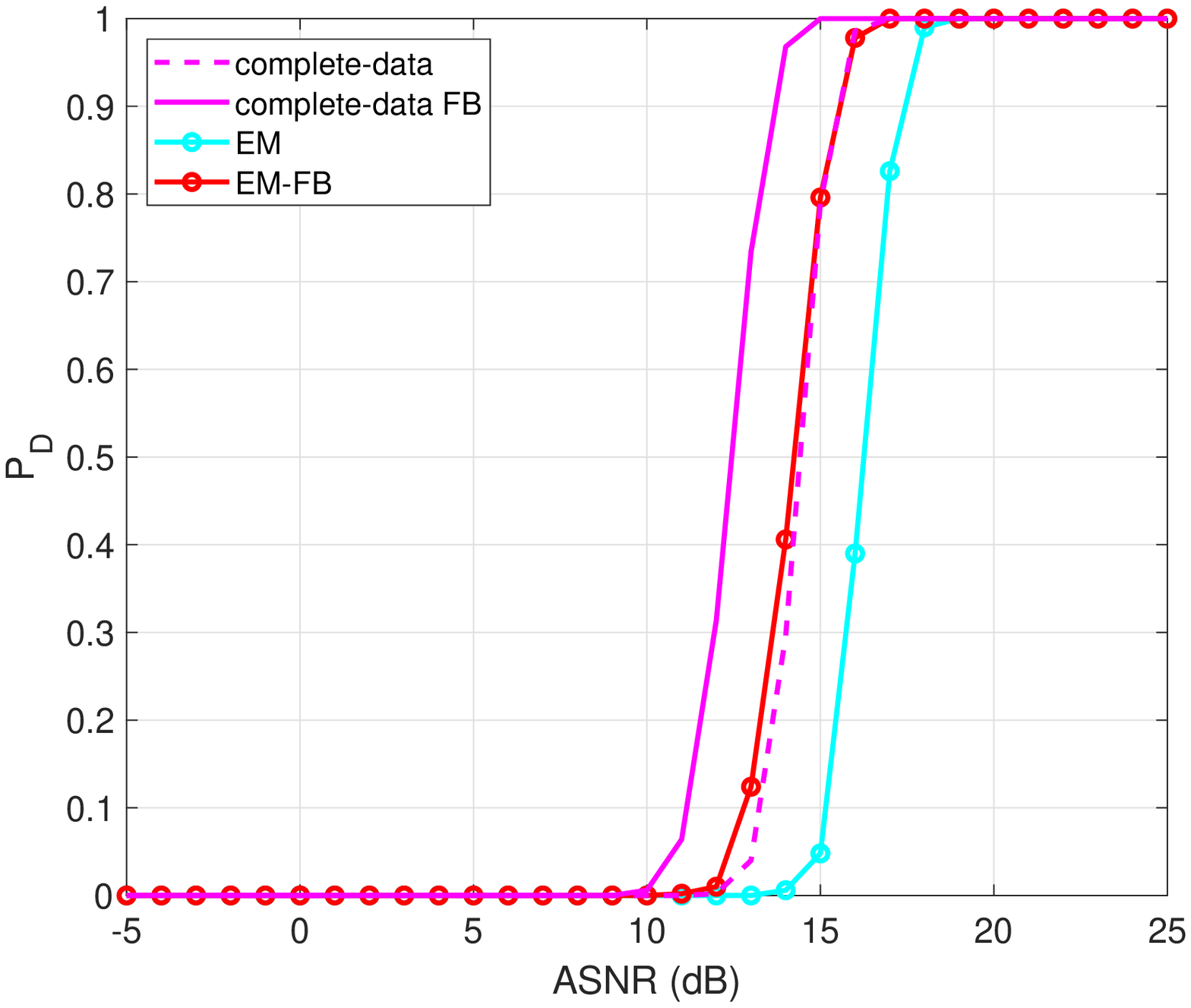}}
       \hspace{10pt}\subfloat[\label{fig:src_detection_comp_pm_0_3_d3_HQC}]{%
        \includegraphics[width=0.32\linewidth]{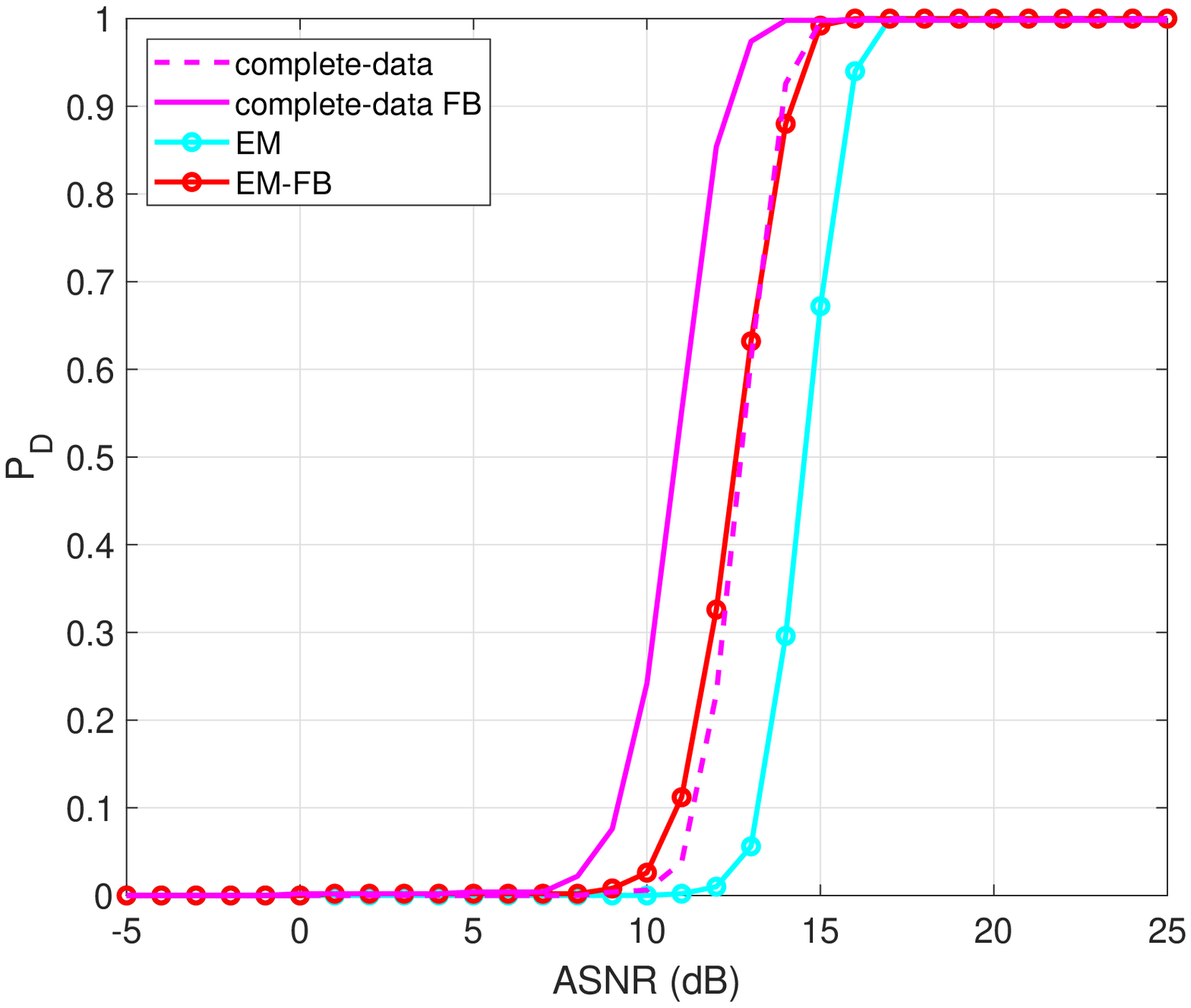}}
   \caption{Comparison of the PD using EM and EM-FB estimation strategies for a ULA with 20 antennas assuming 3 equal-power signals impinging the array with signal separation corresponding to $0.891/N$. Figs. (a), (b), (c) consider $p_{\mathrm m} = 0.1$, whereas $p_{\mathrm m} = 0.3$ is assumed in Figs. (d), (e), (f). Besides, Figs. (a) and (d),  Figs. (b) and (e), and Figs. (c) and (f) consider AIC, MDL, and HQC, respectively.}
  \label{fig:src_detection_comp} 
\end{figure*}

In this section, the performance of the proposed estimation strategy, framed in the context of adaptive beamforming and detection of number of sources, is analyzed.
For both applications it is considered a radar system equipped with a uniform linear array (ULA) pointing in the bore-sight direction ($\theta_0=0$). The array is composed of $N = 20$ antennas with inter-element spacing $d_x = \lambda/2$, where $\lambda$ denotes the radar operating wavelength. Moreover, two different values for the probability $p_{\mathrm m}$ of missing an observation  are considered, i.e., $p_{\mathrm m}=0.1$ or $p_{\mathrm m}=0.3$. For a given $p_{\mathrm m}$, the selection matrix $A_i$ of the $i$-th snapshot is constructed from the diagonal matrix $\bD_i$ whose diagonal entries are statistically IID Bernoulli random variables with parameter $1-p_{\mathrm m}$, skipping rows containing all zeros. Besides, the computation of the observed-data sample covariance matrix $\bS_y = 1/K \sum_{i=i}^{K} \tilde{\by}_i \tilde{\by}_i^\dagger$ is performed employing $\tilde{\by}_i = \bD_i \br_i,\;i=1,\dots,K$.

\subsection{Adaptive Beamforming}
The performance of the adaptive beamformer is analyzed in terms of beampattern shape and normalized average signal-to-interference power ratio (S/I) versus the number of snapshots. Standard Monte Carlo counting techniques over $100$ independent trials to compute the former performance metric and $500$ independent trials for the latter are used. 

In the reported case studies the disturbance covariance matrix is modeled as ${\bM}={\bM}_J+\sigma^2_a{\boldsymbol I}$, where $\sigma^2_a$ is the white noise power level (assumed without loss of generality equal to $0$ dB) and ${\bM}_J$ {is} the jamming covariance contribution. Specifically, denoting by $J_{NB}$ and $J_{WB}$ the number of narrow-band and wide-band jammers {(assumed separated in space)}, ${\bM}_J = \bM_1 + \bM_2$,
where~\cite{farina1992antenna}
\begin{equation}
    \bM_1 = {\sum\limits_{l=1}^{J_{NB}}} \sigma_l^2\: \bv(\theta_l) \bv(\theta_l)^\dagger ,
\end{equation}
with 
\begin{equation}\label{eq:steering_vector}
	\bv(\theta_l) = [1, e^{j \frac{2\pi}{\lambda} d_x \sin(\theta_l)}, \dots, e^{j (N-1)  \frac{2\pi}{\lambda} d_x \sin(\theta_l)}]^{\mathrm{T}} \in \mathbb{C}^N 
\end{equation}
the steering vector in the direction $\theta_l$ of the $l$-th jammer and $\sigma^2_l$ the power of the $l$-th jammer, while
\begin{equation}\label{eq:matrice_wide_band_jammers}
\begin{aligned}
    &\bM_2\left(n,\;m \right) = {\sum\limits_{r=1}^{J_{WB}}} \bar{\sigma}_r^2\: \mbox{sinc}[0.5 {B_f}_r \; (n-m) \zeta_r ]e^{j (n-m) \zeta_r} \;, 
    \end{aligned}
\end{equation}
with $(n,m)\in \{0,\dots,N-1\}^2$ and $\zeta_r = \pi \sin{\theta_r}$; moreover in~(\ref{eq:matrice_wide_band_jammers}), $\bar{\sigma}^2_r$, $\theta_r$, and ${B_f}_r$, represent the power, the DOA, and the fractional bandwidth $B_r/f_0$ (with $B_r$ the actual bandwidth and $f_0$ the carrier frequency) associated with the $r$-th interferer. 
The sinc function appearing in~\eqref{eq:matrice_wide_band_jammers} is defined as $\mbox{sinc}(x) = \sin(x) / (x)$. 

In the following, two different interfering environments are analyzed:
\begin{itemize}
    \item Scenario 1: five narrow-band jammers located at $\theta_l=10+10l$ degrees, $l=1,\dots, 5$ with Jammer to Noise Ratio (JNR) given by $JNR_{l} = 30\text{ dB}$ ($\sigma_l^2={JNR}_l \:\sigma^2_a, \; l=1,\dots, 5$).
    \item Scenario 2: five wide-band jammers $(B_f = 0.03)$ located at $\theta_r=10+10r$ degrees, $r=1,\dots, 5$ with $JNR_{r} = 30\text{ dB}$ ($\sigma_j^2={JNR}_r \:\sigma^2_a, \; r=1,\dots, 5$).
\end{itemize}

The performance of the adaptive beamformer, assuming either $p_{\mathrm m}=0.1$ or $p_{\mathrm m}=0.3$, is analyzed in terms of normalized average S/I in Figs.~\subref*{fig:rob_beam_sc_1_a}, \subref*{fig:rob_beam_sc_1_b}, \subref*{fig:rob_beam_sc_2_a}, and \subref*{fig:rob_beam_sc_2_b}. The resulting beampatterns (assuming $K = 60$), are displayed in Figs.~\subref*{fig:rob_beam_sc_1_c}, \subref*{fig:rob_beam_sc_1_d}, \subref*{fig:rob_beam_sc_2_c}, and \subref*{fig:rob_beam_sc_2_d}. In particular, Figs.~\ref{fig:rob_beam_sc_1} and~\ref{fig:rob_beam_sc_2} refer to the interference environments of Scenario 1 and 2, respectively. 

The proposed strategy employs the EM procedure assuming the uncertainty set  (\ref{set:cm_lb_white_dist_pwr_level}) with the FML computed from $\bS_y$, {used to initialize the EM procedure}.
The beampattern and the normalized average S/I {obtained} using the sample covariance matrix of the complete-data ({as well as its variant based on} FML) and the FML of $\bS_y$, are considered for comparison. As performance benchmark, the clairvoyant beampattern, based on a perfect knowledge of the covariance matrix, is reported too.
A close inspection of the results under the interference environment of Scenario 1 shows that for $p_{\mathrm m} = 0.1$ and $K \ge N$ the performance of the proposed procedure comes closer and closer to the complete-data FML whereas it exhibits for $p_{\mathrm m} = 0.3$ a slight degradation in terms of normalized average S/I in the order of $1$ dB {for $K > N$}, with respect to the {complete-data benchmark}. The effectiveness of the proposed algorithm is also {confirmed by} the more challenging interference environment of Scenario 2, where the performance is very close to the complete-data FML for $p_{\mathrm m} = 0.1$ and experiences a maximum degradation, in terms of normalized avg S/I, lower than $6$ dB, for $p_{\mathrm m} = 0.3$ and $K \ge N$.
{Nevertheless, for} all the configurations, the S/I of the EM-based beampattern approaches the complete-data {performance} as $K$ increases and this represents an indirect proof that the quality of the {proved} covariance estimation procedure improves when more and more snapshots are available for the estimation process.

{As to the} beampattern {analysis}, the inspection of the figures reveals that the EM FML is able to correctly nullify the jammers while preserving low side-lobes levels. 

Finally, Fig.~\ref{fig:sir_comparison} compares the performance of EM FML and EM FML-FB, {highlighting} the capability of FML-FB to benefit from the underlying structure of the covariance matrix.

\subsection{Detection of Number of Sources}
In the following, equal-power signals impinging on the array from different directions $\theta_v$ are considered. The values of the parameters involved in the three analyzed scenarios, each related to a different number of sources, are listed in Table~\ref{tab-parameters}.

\begin{table}[htbp]
	\small
	\centering
	\caption{\label{tab-parameters} Simulation Parameters}
	\begin{tabular}{ccccc}
		\hline
		\hline
		$d$ & $u_1 = \sin(\theta_1)$  & $u_2 = \sin(\theta_2)$ & $u_3 = \sin(\theta_3)$ & $u_4 = \sin(\theta_4)$ \\
		\hline
		$2$ & $-1/2 \;\text{SSBW}$ & $1/2 \;\text{SSBW}$ &  & \\
		$3$ & $-1/2 \;\text{SSBW}$ & $1/2 \;\text{SSBW}$ &  $3/2 \;\text{SSBW}$ &  \\
		$4$ &$-1/2 \;\text{SSBW}$ & $1/2 \;\text{SSBW}$ & $3/2 \;\text{SSBW}$ & $-3/2 \;\text{SSBW}$ \\
		\hline
		\hline
	\end{tabular}
\end{table}

Specifically, $\text{SSBW} = 0.891/N$ denotes the $3$dB single-side beam-width (SSBW) of the considered ULA~\cite{wirth2013radar}, whereas $u_v = \sin(\theta_v)$ is the target angular location of the $v$-th source in the space of directional cosine~\cite{vantrees4}.
Therefore, the covariance matrix is modeled as ${\bM}={\bM}_S+\sigma^2_n{\boldsymbol I}$, where $\sigma^2_n$ is the white noise power level (assumed without loss of generality equal to $0$ dB) and ${\bM}_S$ refers to the useful covariance contribution, given by 
\begin{equation}
	\bM_S = \sigma_s^2 {\sum\limits_{v=1}^d}  \; \bv(\theta_v) \bv(\theta_v)^\dagger ,
\end{equation}
with $\sigma_s^2$ the power of each signal of interest and $\bv(\theta_v)$ is defined as in~\eqref{eq:steering_vector}.

The metric used to assess the detection performance is the Probability of Detection (PD), namely the probability that $\hat{d} = d$~\cite{vantrees4}, {which is} estimated via standard Monte Carlo counting techniques over $500$ independent trials\footnote{Notice that a rank-deficient $\bS_y$, due to a possible selection matrix configuration, is a non-zero probability event. Such realizations are excluded from the Monte Carlo trials.}.
Moreover, the array signal-to-noise ratio (ASNR) is defined as
\begin{equation}
	ASNR = N \frac{\sigma_s^2}{\sigma_n^2} .
\end{equation}
Finally, the detection algorithm {assumes} $K=100$ and a maximum number of sources equal to $N/2 = 10$.

The detection performance is reported in Fig.~\ref{fig:src_detection_1} assuming $p_{\mathrm m} \in \{0.1,\: 0.3\}$ and $K=100$.
In particular, {denoting by $d$ the actual number of sources,} Figs.~\ref{fig:src_detection_1} (a), (b), and (c) {assume $d=2$}, Figs.~\ref{fig:src_detection_1} (d), (e), and (f) $d=3$, whereas Figs.~\ref{fig:src_detection_1} (g), (h), and (i) $d=4$. Moreover, Figs.~\ref{fig:src_detection_1} (a), (d), and (g) refer to AIC, Figs.~\ref{fig:src_detection_1} (b), (e), and (h) consider MDL whereas Figs.~\ref{fig:src_detection_1} (c), (f), and (i) {display} HQC. 

The results highlight that for $p_{\mathrm m}=0.1$ the EM {approach leads to} a performance very close to the complete-data case (with a loss smaller than $1$ dB), and outperforms the basic approach of replacing the missing observations in the complete data with zeros (dashed brown curves), in most of the analyzed case. In fact, a close inspection of the curves shows that only when $d=4$, low ASNR, and with reference to the AIC (Fig.~\ref{fig:src_detection_1} (g)), the basic approach performs better than EM-based technique. This results is not surprising due to the overestimation behavior of the AIC~\cite{vantrees4}.
Besides, the basic strategy may not provide a monotonic behaviour with respect to the ASNR, reflecting the reasonable larger and larger discrepancy between the actual covariance matrix and that heuristically computed.

As expected, the EM-based order selection procedure experiences a performance degradation at $p_{\mathrm m}=0.3$, as compared with the complete-data counterpart. Remarkably, the gap between the EM and the complete-data curves, for $p_{\mathrm m}=0.3$, is less than $3$ dB in the worst case, whereas at the high ASNR regime it is almost absent. As in the case $p_{\mathrm m}=0.1$, EM-based strategy outperforms the basic counterpart, with the only exception of AIC with 4 sources, reported in Fig.~\ref{fig:src_detection_1} (g).

Finally, the detection performance using EM and EM-FB is compared in Fig.~\ref{fig:src_detection_comp}. Inspection of the curves pinpoints that capitalizing on the centro-Hermitian structure, EM-FB achieves higher $P_D$ levels than the unstructured EM in all the considered scenarios, except for the AIC at high ASNR regime where an expected saturation is experienced~\cite{vantrees4}.

\section{Conclusion}\label{section:conclusion}
This paper has considered the problem of structured covariance matrix estimation in the presence of missing data with special attention to a radar signal processing background. After providing a substantial motivation on the study and specifying some constraint sets of particular interest for the covariance matrix, the missing data model is described assuming Gaussian observations. Hence, the ML covariance estimation problem is formulated as the maximization of the observed data log-likelihood. To circumvent the analytical difficulties which are usually connected with the direct optimization of the mentioned function, an iterative maximization procedure based on the EM algorithm is developed and its convergence properties are established. Besides, a closed form expression is computed for the convergence rate.
The theoretical results are capitalized for some specific structural covariance models with reference to two radar applications: adaptive beamforming and detection of the number of sources. General procedures are suggested to construct adaptive beamformers and to detect the number of active sources in a collection of snapshots when missing observations are present. At the analysis stage, extensive numerical results have been discussed to show the effectiveness of the bespoke strategies to handle missing data scenarios. 
In conclusion, the main contributions of the paper can be summarized as followed:
\begin{enumerate}[label=\alph*)]
\item  the development of an EM-based technique for the estimation of a structured covariance matrix in the presence of missing data;
\item the study of the convergence properties for the resulting iterative procedure according to B-stationarity as well as the computation of the rate of convergence;
\item the application of the methodology in the context of two fundamental radar problems: beamforming and detection of the number of sources;
\item the presentation of numerical results aimed at corroborating the theoretical achievements.
\end{enumerate}

Possible future research avenues might include the validation of the approach on real data as well as its application in the context of adaptive target detection \cite{de2015modern} in the presence of missing observations, possibly accounting for compound Gaussian interference. Finally, it is absolutely worth of consideration a careful study on electronic protection techniques when some array elements (or sub-array) of the radar antenna are put in saturation by a strong interference source and, as a consequence, the data can be modeled as missing.

\appendix
\subsection{Proof of~(\ref{eq:C_i})}\label{appendix:proof_Ci}
Following {the same line of reasoning as} in~\cite{10.2307/1165260}{, for any $i \in\{1,\dots, K\}$}
\begin{equation}\label{eq:app1}
	\bC_i = \mathbb{E}[\br_i \br_i^\dagger | \bA_i \br_i,\bM] = \bB_i^{\mathrm{T}} \mathbb{E}[\bB_i \br_i \br_i^\dagger \bB_i^{\mathrm{T}} | \bA_i \br_i,\bM] \bB_i, 
\end{equation}
where
\begin{equation}
	\bB_i = [\bA_i^{\mathrm{T}} \; {\bar{\bA}_i}^{\mathrm{T}} ]^{\mathrm{T}}
\end{equation}
which {satisfies}
\begin{equation}
	\bB_i^{\mathrm{T}} \bB_i = \bI .
\end{equation}
{To provide a closed form expression to $\bC_i$, let us observe that}
\begin{small}
	\begin{equation}
		\begin{aligned}
			&\mathbb{E}[\bB_i \br_i \br_i^\dagger \bB_i^{\mathrm{T}} | \bA_i \br_i, \bM] = \\ & \mathbb{E} \left[ \begin{bmatrix}
				\bA_i \br_i\\ 
				\bar{\bA}_i \br_i
			\end{bmatrix}
			\begin{bmatrix}
				\br_i^\dagger \bA_i^\dagger & \br_i^\dagger \bar{\bA}_i^\dagger
			\end{bmatrix} | \bA_i \br_i, \bM \right] = \\ &  \mathbb{E} \left[ \begin{bmatrix}
				\bA_i \br_i \br_i^\dagger \bA_i^\dagger & \bA_i \br_i \br_i^\dagger \bar{\bA}_i^\dagger \\ 
				\bar{\bA}_i \br_i \br_i^\dagger \bA_i^\dagger & \bar{\bA}_i \br_i \br_i^\dagger \bar{\bA}_i^\dagger
			\end{bmatrix} | \bA_i \br_i, \bM \right] =\\
			& \begin{bmatrix}
				\by_i \by_i^\dagger & \by_i \mathbb{E}\left[\br_i^\dagger \bar{\bA}_i^\dagger | \bA_i \br_i, \bM \right] \\
				\mathbb{E}\left[\bar{\bA}_i \br_i | \bA_i \br_i, \bM \right]\by_i^\dagger & \mathbb{E}\left[\bar{\bA}_i \br_i \br_i^\dagger \bar{\bA}_i^\dagger|\bA_i \br_i, \bM \right]
			\end{bmatrix} = \\ & \begin{bmatrix}
				\by_i \by_i^\dagger & \by_i \bmu_i^\dagger \\
				\bmu_i \by_i^\dagger & \mathbb{E}\left[\bar{\bA}_i \br_i \br_i^\dagger \bar{\bA}_i^\dagger|\bA_i \br_i, \bM \right]
			\end{bmatrix} ,
		\end{aligned}  
	\end{equation}
\end{small}
where
\begin{equation}
	\bmu_i = \mathbb{E}\left[\bar{\bA}_i \br_i | \bA_i \br_i, \bM \right] = \bar{\bA}_i \bM \bA_i^\dagger (\bA_i \bM \bA_i^\dagger)^{-1} \by_i
\end{equation}
and
\begin{equation}
\begin{small}
		\begin{aligned}
		&\mathbb{E}\left[\bar{\bA}_i \br_i \br_i^\dagger \bar{\bA}_i^\dagger|\bA_i \br_i, \bM \right] = \\
		&  \bG_i + \bar{\bA}_i \bM \bA_i^\dagger (\bA_i \bM \bA_i^\dagger)^{-1} \by_i \by_i^\dagger (\bA_i \bM \bA_i^\dagger)^{-1} \bA_i \bM \bar{\bA}_i^\dagger,
	\end{aligned}  
\end{small}
\end{equation}
with
\begin{equation}
	\bG_i = \bar{\bA}_i \bM \bar{\bA}_i^\dagger -  \bar{\bA}_i \bM \bA_i^\dagger (\bA_i \bM \bA_i^\dagger)^{-1} \bA_i \bM \bar{\bA}_i^\dagger.
\end{equation}
Exploiting the above results, {it follows that}{
	\begin{small}
		\begin{equation}
			\begin{aligned}\nonumber
				\bC_i = &  \bB_i^{\mathrm{T}} \mathbb{E}[\bB_i \br_i \br_i^\dagger \bB_i^{\mathrm{T}} | \bA_i \br_i, \bM] \bB_i = \\
				&\bA_i^\dagger \by_i \by_i^\dagger \bA_i +  \bar{\bA}_i^\dagger \bar{\bA}_i \bM \bA_i^\dagger (\bA_i \bM \bA_i^\dagger)^{-1} \by_i \by_i^\dagger \bA_i + \\
				& +\bA_i^\dagger \by_i \by_i^\dagger (\bA_i \bM \bA_i^\dagger)^{-1} \bA_i \bM \bar{\bA}_i^\dagger \bar{\bA}_i + \bar{\bA}_i^\dagger \bar{\bA}_i\bM \bar{\bA}_i^\dagger  \bar{\bA}_i + \\
				&- \bar{\bA}_i^\dagger \bar{\bA}_i \bM \bA_i^\dagger (\bA_i \bM \bA_i^\dagger)^{-1} \bA_i \bM \bar{\bA}_i^\dagger \bar{\bA}_i + \\
				&+  \bar{\bA}_i^\dagger \bar{\bA}_i \bM \bA_i^\dagger (\bA_i \bM \bA_i^\dagger)^{-1} \by_i \by_i^\dagger  (\bA_i \bM \bA_i^\dagger)^{-1}  \bA_i \bM \bar{\bA}_i^\dagger \bar{\bA}_i.
			\end{aligned}
		\end{equation}
	\end{small}
	Finally, after some algebraic manipulations,
	\begin{equation}
		\bC_i =  (\bA_i^\dagger \by_i + \bar{\bA_i}^\dagger \bmu_i) (\bA_i^\dagger \by_i + \bar{\bA_i}^\dagger \bmu_i)^\dagger + \bar{\bA_i}^\dagger \bG_i \bar{\bA_i}
	\end{equation}
	\QEDA
	
	\subsection{Proof of Proposition~\ref{proposition:1}}\label{appendix:proof_prop_1}
	This Appendix is organized in two parts: in Subsection \ref{A_1} the proof of the first item of Proposition \ref{proposition:1} is provided, whereas  Subsection \ref{A_2} deals with second claim.
	\subsubsection{Proof of the first item}\label{A_1} 
	As first step toward the proof, let us observe that a maximizer to Problem (\ref{eq:Mstep}) exists, provided that $\bSigma^{(h-1)}\succ \bf 0 $. Indeed, according to \cite{1456695},
	if  $\bSigma^{(h-1)}\succ \bf 0 $ there exist two positive constants $a$ and $b$ such that Problem (\ref{eq:Mstep}) is equivalent to
	\begin{equation}\label{eq_prob1}
		\left\{ \begin{array}{ll}
			\displaystyle{\max_{\bM}} & -\ln(\det(\bM)) - \tr\{\bM^{-1} \bSigma^{(h-1)}\}\\
			\mbox{s.t.}  & \bM \in \mathcal{C}\\
			&\lambda_{min}(\bM)\geq a\\
			&tr\{\bM\}\leq b
		\end{array} \right.
	\end{equation}
	and any optimal solution $\bar{\btheta}$ to (\ref{eq:Mstep}) must comply with $\lambda_{min}(\bM(\bar{\btheta}))\geq a$ and $tr\{\bM(\bar{\btheta})\}\leq b$. As a result, being $\mathcal{C} \cap \{\bM\succeq {\bf 0}: \lambda_{min}(\bM)\geq a\} \cap \{\bM\succeq {\bf 0}: tr\{\bM\}\leq b \}$ a compact set of positive definite matrices, Problem (\ref{eq_prob1}) admits a global optimal solution $\bM^\star \succ {\bf 0}$, due to Weierstrass Theorem, and any optimal solution is positive definite. Finally, since $\bM^\star \in \mathcal{C}$, there exists $\btheta^\star$ such that $\bM^\star=\bM(\btheta^\star)$, i.e., $\btheta^{(h)}=\btheta^\star$ solves Problem (\ref{eq:Mstep}) and the M-step at the $h$-th iteration is well-defined, being $\hat{\bM}(\btheta^{(h)})\succ \bf 0$. 
	
	Let us now show that 
	\[\bSigma^{(h-1)}=\frac{1}{K}\mathbb{E}[\bD\bD^\dagger| \bY, \{\bA_i\}_{i=1}^N, \hat{\bM}(\btheta^{(h-1)})]\succ \bf 0,\]
	almost surely if $\hat{\bM}(\btheta^{(h-1)})\succ \bf 0$  and $K\geq N$, where $\bD=[\br_1,\ldots,\br_K]\in \mathbb{C}^{N,K}$. To this end, note that, based on Lemma \ref{lemma:1} in Appendix \ref{appendix:proof_lemma}, the random matrix $\bD$ conditioned on $\bY$ is full rank with probability one, for any (but for a zero-measure set) realization of $\bY$. As a consequence, for any $\bv \in \mathbb{C}^N$, the random variable (conditioned on $\bY$)
	\[{\bv^\dagger \left(\bD\bD^\dagger\right)\bv^\dagger}_{| \bY, \{\bA_i\}_{i=1}^N, \hat{\bM}(\btheta^{(h-1)})}\]
	is greater than zero with probability one, which implies
	\[\frac{1}{K}\mathbb{E}[\bD\bD^\dagger| \bY, \{\bA_i\}_{i=1}^N, \hat{\bM}(\btheta^{(h-1)})]\succ {\bf 0}.\]
	Thus, it follows that $\hat{\bM}(\btheta^{(0)})\succ {\bf 0}$ ensures $\hat{\bM}(\btheta^{(h)})\succ \bf 0$ for all $h\geq 1$, namely the M-step is well defined for all $h\geq 1$, because of $$\hat{\bM}(\btheta^{(h-1)})\succ {\bf 0}\Rightarrow \bSigma^{(h-1)} \succ {\bf 0}\Rightarrow \hat{\bM}(\btheta^{(h)})\succ {\bf 0},\, {\forall h\ge1}.$$ Finally, the monotonically increasing behavior of the observed data likelihood function results from the well known properties of EM iterations \cite{10.2307/2984875}. The first item of the proposition is thus proved.
	
	\subsubsection{Proof of the second item}\label{A_2} 
	Before proceeding with the proof, let us introduce the definitions of Bouligand tangent cone and B-stationary point \cite{10.1287/moor.2016.0795, 10.2307/25151818}.
	
	{\bf Definition 1.} Given a set $\mathcal{Z}\subseteq \mathbb{R}^M$,
	the Bouligand tangent cone $\mathcal{T}_{\mathcal{Z}}\left(\bz_0\right)$ of $\mathcal{Z}$ at $\bz_0 \in \mathcal{Z}$, is defined as 
	\begin{small}
		\begin{eqnarray}
			&\mathcal{T}_{\mathcal{Z}}\left(\bz_0\right)=\left\{\bd \in \mathbb{R}^M : \mbox{there exist two sequences }\bz^{(k)}\in \mathcal{Z}\rightarrow \bz_0\,\, \right. \nonumber \\
			& \qquad \qquad \quad \: \mbox{and}\,\, \tau^{(k)}\in\mathbb{R^{++}}\rightarrow 0\,\,\mbox{such that} \left.\bd=\lim_{k \rightarrow \infty} \frac{\bz^{(k)}-\bz_0}{\tau^{(k)}} \right\}\nonumber
		\end{eqnarray}
	\end{small}

	{\bf Definition 2.} Given an optimization problem
	\begin{equation}\label{eq_probaaa}
		\mathcal{P}\left\{ \begin{array}{ll}
			\displaystyle{\max_{\bz}} & f(\bz)\\
			\mbox{s.t.}  & \bz \in \mathcal{Z}\subseteq \mathbb{R}^M\\
		\end{array} \right.
	\end{equation}
	a feasible solution $\bz^\star\in \mathcal{Z}$ is a B-stationary point to Problem $\mathcal{P}$ if
	\begin{align}
		\lim_{\tau_1 \downarrow  0} \frac{f(\bz^{\star}+\tau_1 \bd)-f(\bz^{\star})}{\tau_1}\leq 0,\,\,\,\,\forall \bd \in \mathcal{T}_{\mathcal{Z}}\left(\bz^{\star}\right)
	\end{align}
	In a nutshell, the Bouligand tangent cone generalizes the concept of feasible directions emanating from a point, allowing the extension of the stationarity notion (via the B-stationarity) to a broader class of optimization problems. 
	
	Let us now focus on the convergence of $\mathcal{L}_y(\btheta^{(h)} | \bY, \bA_1, \dots, \bA_K)$ and $\hat{\bM}(\btheta^{(h)})$. In this respect, note that
	\begin{small}
		\begin{equation}
			\begin{aligned}
				&\mathcal{L}_y(\btheta | \bY, \bA_1, \dots, \bA_K) = \\
				& -\sum_{i=1}^{K} \ln(\det(\bA_i {\bM}({\btheta})\bA_i^\dagger ))- \tr\{(\bA_i {\bM}({\btheta})\bA_i^\dagger)^{-1} \by_i \by_i^\dagger\}
				\leq\nonumber\\
				& -\sum_{i=1}^{K} p_i\ln(\lambda_{min}(\bA_i {\bM}({\btheta})\bA_i^\dagger ))-\frac{\|\by_i\|^2}{ \tr\{\bA_i {\bM}({\btheta})\bA_i^\dagger\}}\leq \nonumber\\
				& -\sum_{i=1}^{K} p_i\ln(\lambda_{min}({\bM}({\btheta})))-\frac{\|\by_i\|^2}{ \tr\{{\bM}({\btheta})\}}
			\end{aligned}
		\end{equation}
	\end{small}
	where the last inequality stems from the Eigenvalue Interlacing Theorem \cite{rogerahorn2012}. Hence, being $\mathcal{C}$ a closed set of positive definite matrices, it follows that
	\[\min_{\bM \in \mathcal{C}} \lambda_{min}(\bM)=\delta>0,\]
	which entails
	\begin{small}
		\begin{equation}\label{bound}
			\mathcal{L}_y(\btheta | \bY, \bA_1, \dots, \bA_K)\leq -\left(\sum_{i=1}^{K} p_i\right)\log(\delta)- \frac{\displaystyle{\sum_{i=1}^K}\|\by_i\|^2}{\tr\{\bM(\theta))\}}
		\end{equation}
	\end{small}
	As a result, the log-likelihood diverges to $-\infty$ as $\tr\{\bM\}\rightarrow +\infty$ and, following the same line of reasoning as that leading to the equivalent formulation of the M-step in (\ref{eq_prob1}), an optimal solution $\hat{\bM}(\hat{\btheta}_{ML})\succ \bf 0$ to Problem (\ref{eq:problem_theta}) exists. Now, since 
	$\mathcal{L}_y({\bM}(\btheta^{(h)}) | \bY, \bA_1, \dots, \bA_K)$
	defines an increasing sequence of values
	\begin{enumerate}
		\item $\mathcal{L}_y(\btheta^{(h)} | \bY, \bA_1, \dots, \bA_K)$ converges to a finite value, as it is bounded above by $\mathcal{L}_y(\hat{\bM}(\hat{\btheta}_{ML}) | \bY, \bA_1, \dots, \bA_K)$
		\item ${\bM}(\btheta^{(h)}),\,h\geq 1$ is a bounded sequence, due to (\ref{bound}) and {$\|\bM\|_F \leq {\tr{\left(\bM\right)}}$}.
	\end{enumerate}
	
	Let now us assess the convergence properties of $\btheta^{(h)}$, $h\geq 1$. To proceed further,  observe that $\btheta^{(h)}$, $h\geq 1$, is a bounded sequence, being $\bM(\btheta)$ a norm coercive mapping\footnote{A mapping $\bM(\cdot): \bx \in \mathbb{R}^{M_2}\rightarrow \bM(\bx)\in \mathbb{C}^{M_1,M_1}$ is said norm coercive if $\|\bx\|\rightarrow\infty$ implies {$\|\bM(\bx)\|_F\rightarrow\infty$}.}. Hence, let $\btheta^{\star}$ be a limit point of $\btheta^{(h)}$, $h\geq 1$, whose existence is ensured by the  boundedness of $\btheta^{(h)}$, $h\geq 1$. Owing to the interpretation of the EM procedure as a minorization-maximization optimization technique~\cite{doi:10.1198/0003130042836}, it can be shown that $\btheta^\star$ is a global optimal solution to
	\begin{equation}\label{eq:Mstep_1}
		\operatorname*{arg max}\limits_{\btheta:\bM(\btheta) \in \mathcal{C}} Q(\btheta, \btheta^{\star})
	\end{equation}
	Indeed, denoting by $g(\btheta)=\mathcal{L}_y(\btheta | \bY, \bA_1, \dots, \bA_K)-Q(\btheta, \btheta)$, it follows that
	\begin{align}
		\mathcal{L}_y(\btheta | \bY, \bA_1, \dots, \bA_K) \;=\;&Q(\btheta, \btheta)+g(\btheta),\,\forall \btheta \nonumber\\
		\mathcal{L}_y(\btheta | \bY, \bA_1, \dots, \bA_K) \;\geq\;&Q(\btheta, \btheta_1)+g(\btheta_1),\,\forall \btheta, \forall \btheta_1\nonumber
	\end{align}
	and
	\begin{eqnarray}
		\begin{aligned}
				&Q\left(\btheta, \btheta^{\left(i_{(h)}\right)}\right)+g\left(\btheta^{\left(i_{(h)}\right)}\right) \\ &\leq Q\left(\btheta^{\left(i_{(h)}+1\right)}, \btheta^{\left(i_{(h)}\right)}\right)+ g\left(\btheta^{\left(i_{(h)}\right)}\right)\\
				&\leq \mathcal{L}_y\left(\btheta^{\left(i_{(h)}+1\right)} | \bY, \bA_1, \dots, \bA_K\right)\\
				&\leq \mathcal{L}_y\left(\btheta^{\left(i_{(h+1)}\right)} | \bY, \bA_1, \dots, \bA_K\right)\\
				&=  Q\left(\btheta^{\left(i_{(h+1)}\right)}, \btheta^{\left(i_{(h+1)}\right)}\right)+ g\left(\btheta^{\left(i_{(h+1)}\right)}\right)
		\end{aligned}
	\end{eqnarray}
	where $i_{(h)}$ indexes the subsequence extracted by $\btheta^{(h)}$ that converges to $\btheta^\star$, i.e., $\btheta^{(i_{(h)})}\rightarrow \btheta^\star$, as $h\rightarrow \infty$. Now, since $Q(\btheta, \btheta_1)+g(\btheta_1)$ is a continuous function with respect to $(\btheta, \btheta_1)$, 
	\[Q(\btheta, \btheta^{\star})+g(\btheta^{\star})\leq Q(\btheta^{\star}, \btheta^{\star})+g(\btheta^{\star}) . \]
	
	As a consequence, since $\btheta^{\star}$ is a global optimal solution to Problem  (\ref{eq:Mstep_1}) and  $Q(\btheta, \btheta^{\star})+g(\btheta^{\star})$ is a differentiable function with respect to $\btheta$ {in a neighborhood of $\btheta^\star$} (being it the composition of differentiable functions and $\bM(\btheta^\star)\succ \bf 0$), it follows that $\btheta^{\star}$ is a B-stationary point to Problem (\ref{eq:Mstep_1}) \cite{10.1287/moor.2016.0795, 10.2307/25151818}, namely
	\begin{align}
		\lim_{\tau \downarrow  0} \frac{Q(\btheta^{\star}+\tau \bd, \btheta^{\star})-Q(\btheta^{\star}, \btheta^{\star})}{\tau}\leq 0,\,\,\,\,\forall \bd \in \mathcal{T}_{\mathcal{F}}\left(\btheta^{\star}\right)
	\end{align}
	where $\mathcal{T}_{\mathcal{F}}\left(\btheta^{\star}\right)$ is the Bouligand tangent cone of $\mathcal{F}=\left\{\btheta:\,\btheta \in \mathcal{C}\right\}$ at $\btheta^{\star}$.
	
	Finally, being $\mathcal{L}_y(\btheta | \bY, \bA_1, \dots, \bA_K)$
	a differentiable function in a neighborhood of $\btheta^\star$, leveraging Lemma 2 in \cite{8454321} it holds true
	\begin{eqnarray}
		\begin{small}\nonumber
			\begin{aligned}
				& \lim\limits_{\tau \downarrow  0} \frac{\mathcal{L}_y(\btheta^{\star}+\tau \bd | \bY, \bA_1, \dots, \bA_K)-\mathcal{L}_y(\btheta^{\star}| \bY, \bA_1, \dots, \bA_K)}{\tau}= \\
				&\lim\limits_{\tau \downarrow  0} \frac{Q(\btheta^{\star}+\tau \bd, \btheta^{\star})-Q(\btheta^{\star}, \btheta^{\star})}{\tau} , \;\; \forall \bd, &\; 
			\end{aligned}
		\end{small}
	\end{eqnarray}
	which implies that $\btheta^{\star}$ is a B-stationary point to Problem (\ref{eq:problem_theta}).
	
	\subsection{Lemma \ref{lemma:1} and its proof}\label{appendix:proof_lemma}
	\begin{lemma}\label{lemma:1}
		Let $\bX\in \mathbb{C}^{N \times K}$ be a random matrix whose columns are IID zero-mean circularly symmetric Gaussian random vectors with covariance matrix $\bar{\bM}\succ \bf 0$ and $K\geq N$. Then, denoting by $\bz\in \mathbb{C}^{N_1}$ a vector composed by $N_1\leq N K$ distinct entries of $\bX$, \begin{equation}
			{\mathbb P}\left(\det\left({\bf X} {\bf X}^\dagger \right)> 0| \bz, \bar{\bM}\right)=1
		\end{equation}
		for any $\bz\in \mathbb{C}^{N_1} \setminus \Omega_0$, with $\Omega_0\subseteq \mathbb{C}^{N_1}$ a zero-measure set. 
	\end{lemma}
	
	\begin{IEEEproof}
		According to \cite{james1964,7891546}
		\begin{eqnarray}\nonumber
			1 = {\mathbb{P}}\left(\det\left(\bX\bX^\dagger\right)>0|\bar{\bM}\right) =  \mathbb{E}[{\mathbb{P}}\left(\det\left({\bf X} {\bf X}^\dagger\right)>0| \bz, \bar{\bM}\right)]
		\end{eqnarray}
		where the second equality stems from the Fubini-Tonelli Theorem \cite{billingsley2012probability}.
		As consequence, 
		\begin{small}
			\begin{eqnarray}\label{conditoned}
				\begin{aligned}
					{\mathbb{P}}\left(\det\left(\bX\bX^\dagger\right)>0| \bz,\bar{\bM}\right)&=1,\, \mbox{almost surely, i.e.,}\, \forall \bz,  \mbox{but}\\
					&\,\,\,\,\,\, \mbox{for a zero-probability event}
				\end{aligned}
			\end{eqnarray}
		\end{small}
		Finally, being $\bz$ a Gaussian random vector with a positive definite covariance matrix, (\ref{conditoned}) is equivalent to the existence of a zero-measure set $\Omega_0\subseteq \mathbb{C}^{N_1}$, such that
		\[{\mathbb{P}}\left(\det\left(\bX\bX^\dagger\right)>0| \bz,\bar{\bM}\right)=1, \,\,\forall \bz \in \mathbb{C}^{N_1}\setminus \Omega_0.\]
	\end{IEEEproof}

	\subsection{Derivation of \eqref{eq:fobs} and \eqref{eq:fem}}
	In the following, {closed-form expression} of $\bF_{EM}$ and $\bF_{obs}$ are derived {with reference to \eqref{eq:observed_log_likelihood} and \eqref{eq:complete_log_likelihood}}. In this respect, note that the $(l,m)$-th element of $\bF_{EM} = \left. \mathbb{E}\left[ - \nabla_{\btheta} \nabla_{\btheta}^{\mathrm{T}}  \mathcal{L}_r(\btheta) | \bY, \btheta \right] \right|_{\btheta = {\hat{\btheta}_{ML}}}$ is given by
	\begin{equation}\label{eq:defFEM}
		\begin{aligned}
			\bF_{EM}(l, m) = \left. \mathbb{E}\left[ - \frac{\partial^2 \mathcal{L}_r(\btheta)}{\partial \theta_l \partial \theta_m}  | \bY, \btheta \right] \right|_{\btheta = {\hat{\btheta}_{ML}}},
		\end{aligned}
	\end{equation}
	where $(l,m)\in \{1,\dots,V\}^2$ and $\mathcal{L}_r(\btheta)$ is the complete-data log-likelihood given by~\eqref{eq:complete_log_likelihood}.
	Furthermore, 
	\begin{equation}
		{  - \frac{\partial^2 \mathcal{L}_r(\btheta)}{\partial \theta_l \partial \theta_m} }= K \frac{\partial^2 \ln(\det(\bM(\btheta)))}{\partial \theta_l \partial \theta_m} + K \frac{\partial^2 \tr\left\{ \bM(\btheta)^{-1}\bS \right\}}{\partial \theta_l \partial \theta_m}, 
	\end{equation}
	{with $ \bS$ given in~\eqref{eq:sample_covariance}, whereas}~\cite[A.393]{vantrees4}
	\begin{small}
		\begin{eqnarray}
			\begin{aligned}
				&\frac{\partial^2 \ln(\det(\bM(\btheta)))}{\partial \theta_l \partial \theta_m} = \\
				& \tr\left\{ - \bM(\btheta)^{-1} \frac{\partial \bM(\btheta)}{\partial \theta_l} \bM(\btheta)^{-1}  \frac{\partial \bM(\btheta)}{\partial \theta_m} + \bM(\btheta)^{-1} \frac{\partial^2 \bM(\btheta)}{\partial \theta_l \partial \theta_m}  \right\}
			\end{aligned}
		\end{eqnarray}
	\end{small}
	and~\cite[A.391]{vantrees4}, \cite[A.392]{vantrees4}
	\begin{small}
		\begin{equation}
			\begin{aligned}
				&\frac{\partial^2  \tr\left\{ \bM(\btheta)^{-1}\bS \right\}}{\partial \theta_l \partial \theta_m} = \\
				&  { \frac{\partial}{\partial \theta_l}} \left[ {-\tr} \left\{  \bM(\btheta)^{-1}  \bS  \bM(\btheta)^{-1} \frac{\partial \bM(\btheta)}{\partial \theta_m} \right\}\right]  =\\ 
				=&-\tr \left\{  \bM(\btheta)^{-1}  \bS  \bM(\btheta)^{-1} \frac{\partial^2 \bM(\btheta)}{\partial \theta_l \partial \theta_m}   +  \right. \\
				&\left. \frac{\partial \bM(\btheta)^{-1}}{\partial \theta_l} \bS \bM(\btheta)^{-1}  \frac{\partial \bM(\btheta)}{\partial \theta_m} + \right.\\
				&\left. \bM(\btheta)^{-1} \bS \frac{\partial \bM(\btheta)^{-1}}{\partial \theta_l} \frac{\partial \bM(\btheta)}{\partial \theta_m}\right\} = \\
				& = -\tr\left\{ \bM(\btheta)^{-1} \bS  \bM(\btheta)^{-1} \frac{\partial^2 \bM(\btheta)}{\partial \theta_l \partial \theta_m} - \right. \\
				&\left.	\bM(\btheta)^{-1} \frac{\partial \bM(\btheta)}{\partial \theta_l}  \bM(\btheta)^{-1} \bS  \bM(\btheta)^{-1} \frac{\partial \bM(\btheta)}{\partial \theta_m} + \right. \\
				&\left. - \bM(\btheta)^{-1} \bS  \bM(\btheta)^{-1}	\frac{\partial \bM(\btheta)}{\partial \theta_l} \bM(\btheta)^{-1} 	\frac{\partial \bM(\btheta)}{\partial \theta_m}\right\}.
			\end{aligned}
		\end{equation}
	\end{small}

	Therefore, 
	\begin{eqnarray} \label{eq:fem_not_expect}
		\begin{small}
			\begin{aligned}\nonumber
				&- \frac{\partial^2 \mathcal{L}_r(\btheta)}{\partial \theta_l \partial \theta_m}  = \\ &
				K \tr\left\{ - \bM(\btheta)^{-1} \frac{\partial \bM(\btheta)}{\partial \theta_l} \bM(\btheta)^{-1}  \frac{\partial \bM(\btheta)}{\partial \theta_m} + \bM(\btheta)^{-1} \frac{\partial^2 \bM(\btheta)}{\partial \theta_l \partial \theta_m}  \right\} + \\
				& + K \tr\left\{  \left[ \bM(\btheta)^{-1} \frac{\partial \bM(\btheta)}{\partial \theta_l}  \bM(\btheta)^{-1}  	\frac{\partial \bM(\btheta)}{\partial \theta_m} -\bM(\btheta)^{-1}   \frac{\partial^2 \bM(\btheta)}{\partial \theta_l \partial \theta_m} + \right. \right. \\
				& \left. \left. \qquad \qquad +  \bM(\btheta)^{-1} \frac{\partial \bM(\btheta)}{\partial \theta_m} \bM(\btheta)^{-1} \frac{\partial \bM(\btheta)}{\partial \theta_l} \right] \bM(\btheta)^{-1} \bS \right\}
			\end{aligned}
		\end{small}
	\end{eqnarray}
	Finally, substituting the above equation into \eqref{eq:defFEM} leads to
	\begin{equation}
		\begin{small}
			\begin{aligned}
				&	\bF_{EM}(l, m) = \\ &   K \tr\left\{ - \bM({\hat{\btheta}_{ML}})^{-1} \frac{\partial \bM(\btheta)}{\partial \theta_l} \bM({\hat{\btheta}_{ML}})^{-1}  \frac{\partial \bM(\btheta)}{\partial \theta_m} + \right. \\ & \left. \bM({\hat{\btheta}_{ML}})^{-1} \frac{\partial^2 \bM(\btheta)}{\partial \theta_l \partial \theta_m}  \right\} + \\
				& + K \tr\left\{  \left[ \bM({\hat{\btheta}_{ML}})^{-1} \frac{\partial \bM(\btheta)}{\partial \theta_l} \right. \right. \\ & \left. \left.  \bM({\hat{\btheta}_{ML}})^{-1}  	\frac{\partial \bM({\hat{\btheta}_{ML}})}{\partial \theta_m} -\bM({\hat{\btheta}_{ML}})^{-1}   \frac{\partial^2 \bM(\btheta)}{\partial \theta_l \partial \theta_m} + \right. \right. \\
				& \left. \left. \qquad \qquad +  \bM({\hat{\btheta}_{ML}})^{-1} \frac{\partial \bM(\btheta)}{\partial \theta_m} \bM({\hat{\btheta}_{ML}})^{-1} \frac{\partial \bM(\btheta)}{\partial \theta_l} \right] \right. \\ & \left.  \bM({\hat{\btheta}_{ML}})^{-1} \bSigma^* \right\} , \quad (l,m)\in \{1,\dots,V\}^2
			\end{aligned}
		\end{small}
	\end{equation}
	with
	\begin{equation}
		\bSigma^* = \frac{1}{K} \sum_{i=1}^K \mathbb{E}[\br_i \br_i^\dagger | \by_i, \bA_i, \hat{\bM}({\hat{\btheta}_{ML}})].
	\end{equation}
	
	Let us now derive the expression of $\bF_{obs}$.
	The $(l,m)$-th element of $\bF_{obs} = \left. - \nabla_{\btheta} \nabla_{\btheta}^{\mathrm{T}}  \mathcal{L}_y(\btheta) \right|_{\btheta = {\hat{\btheta}_{ML}}}$ is given by
	\begin{small}
		{\begin{equation}\label{eq:def_F_obs}
				\begin{aligned}
					&\bF_{obs}(l, m) = \left. - \frac{\partial^2 \mathcal{L}_r(\btheta)}{\partial \theta_l \partial \theta_m} \right|_{\btheta = {\hat{\btheta}_{ML}}} = \\& \sum_{i=1}^K \frac{\partial^2 \ln(\det(\bM_i({\hat{\btheta}_{ML}})))}{\partial \theta_l \partial \theta_m} + \frac{\partial^2 \tr \left\{\bM_i({\hat{\btheta}_{ML}})^{-1} \by_i\by_i^\dagger \right\}}{\partial \theta_l \partial \theta_m} , \\ & \quad (l,m)\in \{1,\dots,V\}^2
				\end{aligned}
		\end{equation}}
	\end{small}
	
	with~\cite[A.393]{vantrees4}
	\begin{eqnarray}
		\begin{aligned}
			&{\frac{\partial^2 \ln(\det(\bM_i({\hat{\btheta}_{ML}})))}{\partial \theta_l \partial \theta_m}}  =  \\
			&\tr \left\{ -\bM_i({\hat{\btheta}_{ML}})^{-1} \frac{\partial \bM_i(\btheta)}{\partial \theta_l}   \bM_i({\hat{\btheta}_{ML}})^{-1} \frac{\partial \bM_i(\btheta)}{\partial \theta_m}   + \right. \\ & \left. \bM_i({\hat{\btheta}_{ML}})^{-1}\frac{\partial^2 \bM_i(\btheta)}{\partial \theta_l \partial \theta_m}  \right\} 	
		\end{aligned}
	\end{eqnarray}
	and~\cite[A.391]{vantrees4}, \cite[A.392]{vantrees4}
	\begin{small}
		\begin{equation}
			\begin{aligned}\nonumber
				&{\frac{\partial^2 \tr \left\{\bM_i({\hat{\btheta}_{ML}})^{-1} \by_i\by_i^\dagger \right\}}{\partial \theta_l \partial \theta_m}}  = \\
				& \frac{\partial}{\partial \theta_l}  \left[ -\tr \left\{ \bM_i({\hat{\btheta}_{ML}})^{-1} \by_i \by_i^\dagger \bM_i({\hat{\btheta}_{ML}})^{-1}  \frac{\partial \bM_i(\btheta)}{\partial \theta_m}   \right\}   \right] = \\
				&= -\tr \left\{ \bM_i({\hat{\btheta}_{ML}})^{-1} \by_i \by_i^\dagger \bM_i({\hat{\btheta}_{ML}})^{-1}  \frac{\partial^2 \bM_i(\btheta)}{\partial \theta_l \partial \theta_m} + \right. \\
				&   - \bM_i({\hat{\btheta}_{ML}})^{-1} \frac{\partial \bM_i(\btheta)}{\partial \theta_l} \bM_i({\hat{\btheta}_{ML}})^{-1} \by_i \by_i^\dagger \bM_i({\hat{\btheta}_{ML}})^{-1} \frac{\partial \bM_i(\btheta)}{\partial \theta_m} + \\
				&\left.   -\bM_i({\hat{\btheta}_{ML}})^{-1} \by_i \by_i^\dagger \bM_i({\hat{\btheta}_{ML}})^{-1} \frac{\partial \bM_i(\btheta)}{\partial \theta_l}  \bM_i({\hat{\btheta}_{ML}})^{-1} \frac{\partial \bM_i(\btheta)}{\partial \theta_m}   \right\} .
			\end{aligned}
		\end{equation}
	\end{small}
	Exploiting the above results, the expression of $\bF_{obs}(l, m)$ is given in \eqref{eq:Fobs}, with $(l,m)\in \{1,\dots,V\}^2$ and $\bM_i({\hat{\btheta}_{ML}}) = \bA_i \bM({\hat{\btheta}_{ML}}) \bA_i^\dagger$.

	\newcounter{eq3}
	\begin{figure*}[htb]
		\hrulefill 
		\normalsize
		\begin{equation}\label{eq:Fobs}
			\begin{small}
				\begin{aligned}
					\bF_{obs}(l, m) = \sum_{i=1}^K & \tr\left\{-\bM_i({\hat{\btheta}_{ML}})^{-1} \frac{\partial \bM_i(\btheta)}{\partial \theta_l}  \bM_i({\hat{\btheta}_{ML}})^{-1}   \frac{\partial \bM_i(\btheta)}{\partial \theta_m} +  \bM_i({\hat{\btheta}_{ML}})^{-1}   \frac{\partial^2 \bM_i(\btheta)}{\partial \theta_l \partial \theta_m}\right\} + \\
					& -\tr \left\{ \bM_i({\hat{\btheta}_{ML}})^{-1} \by_i \by_i^\dagger \bM_i({\hat{\btheta}_{ML}})^{-1}  \frac{\partial^2 \bM_i(\btheta)}{\partial \theta_l \partial \theta_m} + \right. \\
					& \left. - \bM_i({\hat{\btheta}_{ML}})^{-1} \frac{\partial \bM_i(\btheta)}{\partial \theta_l} \bM_i({\hat{\btheta}_{ML}})^{-1} \by_i \by_i^\dagger \bM_i({\hat{\btheta}_{ML}})^{-1} \frac{\partial \bM_i(\btheta)}{\partial \theta_m} +   \right. \\
					& { \left. \qquad \quad -\bM_i({\hat{\btheta}_{ML}})^{-1} \by_i \by_i^\dagger \bM_i({\hat{\btheta}_{ML}})^{-1} \frac{\partial \bM_i(\btheta)}{\partial \theta_l}  \bM_i({\hat{\btheta}_{ML}})^{-1} \frac{\partial \bM_i(\btheta)}{\partial \theta_m}   \right\} }
				\end{aligned}	
			\end{small}
		\end{equation}
		\hrulefill 
	\end{figure*}

\bibliographystyle{IEEEbib}
\bibliography{IEEEabrv,references}

\end{document}

%% file: Def.tex
\usepackage{amsmath}
\usepackage{amsfonts}
\usepackage{mathrsfs}
\usepackage{pifont}
\usepackage{amssymb}
\usepackage{verbatim}
\usepackage{upgreek}
\usepackage[final]{graphicx}
\usepackage{booktabs}
\usepackage{bm}
\usepackage{setspace}
\usepackage{url}
\usepackage[utf8]{inputenc}
\usepackage{algorithmic}
\usepackage{algorithm}
\usepackage[overload]{empheq}
\usepackage{fancyhdr}
\usepackage{cite}

\usepackage{xcolor}

\usepackage{amsthm}

\ifCLASSOPTIONcompsoc
    \usepackage[caption=false, font=normalsize, labelfont=sf, textfont=sf, subrefformat=parens,labelformat=parens]{subfig}
\else
\usepackage[caption=false, font=footnotesize, subrefformat=parens,labelformat=parens]{subfig}
\fi

\usepackage{etoolbox}
\robustify\subref

\usepackage{graphicx}
\usepackage{grffile}
\usepackage[shortlabels]{enumitem}

\newcommand*{\QEDA}{\hfill\ensuremath{\blacksquare}}%

\newcommand{\diag}{\textbf{diag}}

\newcommand{\tr}{\text{tr}}

\newcommand{\bSigma}{\boldsymbol{\Sigma}}
\newcommand{\bLambda}{\boldsymbol{\Lambda}}

\newcommand{\ba}{\boldsymbol{a}}

\newcommand{\bd}{\boldsymbol{d}}
\newcommand{\be}{\boldsymbol{e}}

\newcommand{\bn}{\boldsymbol{n}}

\newcommand{\br}{\boldsymbol{r}}
\newcommand{\bs}{\boldsymbol{s}}

\newcommand{\bv}{\boldsymbol{v}}
\newcommand{\bw}{\boldsymbol{w}}
\newcommand{\bx}{\boldsymbol{x}}
\newcommand{\by}{\boldsymbol{y}}
\newcommand{\bz}{\boldsymbol{z}}

\newcommand{\bA}{\boldsymbol{A}}
\newcommand{\bB}{\boldsymbol{B}}
\newcommand{\bC}{\boldsymbol{C}}
\newcommand{\bD}{\boldsymbol{D}}

\newcommand{\bF}{\boldsymbol{F}}
\newcommand{\bG}{\boldsymbol{G}}

\newcommand{\bI}{\boldsymbol{I}}
\newcommand{\bJ}{\boldsymbol{J}}

\newcommand{\bM}{\boldsymbol{M}}

\newcommand{\bR}{\boldsymbol{R}}
\newcommand{\bS}{\boldsymbol{S}}

\newcommand{\bU}{\boldsymbol{U}}
\newcommand{\bV}{\boldsymbol{V}}

\newcommand{\bX}{\boldsymbol{X}}
\newcommand{\bY}{\boldsymbol{Y}}

\newcommand{\bmu}{\boldsymbol{\mu}}
\newcommand{\btheta}{\boldsymbol{\theta}}
\newcommand{\bPhi}{\boldsymbol{\Phi}}

\newcommand{\bzero}{\boldsymbol{0}}
\newcommand{\bone}{\boldsymbol{1}}

\newtheorem{theorem}{Theorem}[section]

\newtheorem{lemma}[theorem]{Lemma}